\documentclass[a4paper,11pt]{article}
\pdfoutput=1
\usepackage{epsfig}
\usepackage{graphicx}

\makeatletter

\@addtoreset{equation}{section}
\makeatother
\newcommand{\eq}{\begin{equation}}
\newcommand{\eqx}{\end{equation}}
\newcommand{\eqs}{\begin{equation*}}
\newcommand{\eqsx}{\end{equation*}}
\newcommand{\eqn}{\begin{eqnarray}}
\newcommand{\eqnx}{\end{eqnarray}}
\newcommand{\alg}{\begin{align}}
\newcommand{\algx}{\end{align}}

\newcommand{\f}[2]{\frac{#1}{#2}}

\begin{document}

\begin{titlepage}
\vskip1cm
\begin{flushright}
UOSTP {\tt 130901}
\end{flushright}
\vskip 2.25cm
\centerline{\Large
\bf Multi-faced Black Janus  and Entanglement
}
\vskip 1.25cm \centerline{  Dongsu Bak, and  Hyunsoo Min }
\vspace{1cm} \centerline{  Physics Department,
University of Seoul, { Seoul 130-743, Korea}} 
\vskip 0.75cm \centerline{(\tt dsbak@uos.ac.kr, hsmin@dirac.uos.ac.kr)} 
\vspace{2.5cm}
\centerline{\bf Abstract} \vspace{0.75cm} \noindent
To the second order of the deformation parameter, we construct  the black brane solutions, 
which are dual to the multiple interface deformations of  conformal field theories. 
We compute their thermodynamic quantities from the gravity and the field theory 
sides, and find a precise 
agreement, confirming the validity 
of the AdS/CFT correspondence.
The correlation of two separate interfaces induces a Casimir energy and an interesting form of 
correlated entropy contribution. We comment on the properties of the interface lattice system too.

\vspace{1.75cm}

\end{titlepage}

\section{Introduction}
The Janus solutions \cite{Bak:2003jk} in various supergravity theories, which  
are dual to the interface conformal field theories (ICFT), provide
well-controlled deformations of the AdS/CFT correspondence \cite{Maldacena:1997re,Gubser:1998bc,Witten:1998qj}.
In the simplest examples, the bulk gravity is described by the Einstein-scalar theory with a negative cosmological 
constant. One uses the AdS$_d$ slicing of the AdS$_{d+1}$ and makes the massless scalar field dependent on the slicing 
coordinate. The scalar field approaches two different constant values at the boundary of the space, which results in a jump of 
the coupling through the interface of the boundary CFT. Out of the original SO$(d,2)$ conformal symmetries of the CFT, this simple 
interface
deformation 
preserves  one lower dimensional conformal symmetries, SO$(d-1,2)$. See Refs.~\cite{D'Hoker:2007xy}-\cite{Clark:2013mfa}
for further development and other related studies.

Recently in Ref.~\cite{bgj}, the black Janus solution 
in three dimensions was considered, which is the Janus deformation 
of the well known BTZ black hole solution \cite{Banados:1992wn}. The corresponding boundary theory is given by 
 the simple interface deformation of the CFT on ${R}^{1,1}$. Strictly speaking, the black Janus geometry
is dual to the finite-temperature version of the ICFT. It was found that 
the interface system 
carries a zero-temperature entropy whose exponential counts
the degeneracy of the ground states. 
Using the solution, the thermodynamic quantities of this ICFT can be calculated based on the dictionary
of the AdS/CFT correspondence.  In the field theory side,  these results 
were reproduced, 
via the so-called conformal perturbation theory, to the second order of the deformation parameter. 

In this paper, we would like to extend the above analysis to the general multiple interface system. We shall 
approach the problem by a perturbative construction of solutions. The zeroth order of the perturbation theory 
begins with the BTZ black hole solution. Our background is dependent upon a single (temperature) scale reflecting 
the conformal invariance of the underlying system. Around the BTZ background, we shall consider general 
static perturbation of the Einstein-scalar theory. The scalar perturbation begins with the term of  $O(\gamma)$
where $\gamma$ is our deformation  parameter counting the order of the perturbation series. The order of the general scalar 
perturbation terms is counted as $\gamma^{2n+1}$ with $n\ge 0$. On the other hand, the perturbation in the geometric 
part begins with terms of $O(\gamma^2)$ and the order of its general perturbation  is counted as $\gamma^{2n+2}$ with 
$n \ge 0$. 

We shall first solve the leading order term of the scalar part, which satisfies a linear partial differential equation.
 One can show that the most general boundary condition for 
the scalar term is that its values should approach piecewise constants at the boundary of the space. For a single interface,
the solution of the equation was already constructed in \cite{bgj}. The general solution can be given by the linear superposition of 
each single-interface solution whose center is located at an arbitrary position of the boundary spatial direction. The leading order 
of  the geometric part is reduced to  linear differential equations with source terms determined 
by the above solution of the scalar part. The source is  consisting of two different kinds of terms. One is from
 the diagonal contributions of  each scalar solution for each  interface. The solution with  a source term of this kind 
is known in Ref.~\cite{bgj}. The other kind is consisting of the off-diagonal contributions  of a set of scalar solutions from two different 
interfaces.

Up to the translational freedom in the boundary spatial direction, the source term and the corresponding 
equations are characterized by the separation distance $l$ of the two interfaces.  The existence of the extra scale $l$ breaks the
underlying conformal invariance in a rather clear manner.  Hence solving this geometric part becomes highly nontrivial.
However, it should be 
in principle possible to find solutions since we are dealing with linear partial differential equations with specified source 
terms. 
To solve those nontrivial partial differential equations, we adopt a suitable ansatz even though its structure is not completely justified. 
We shall find explicit forms of solutions 
(with a freedom of adding homogeneous solutions), which satisfy all of the required physical
properties. Since the procedure of finding solutions is nontrivial, we shall present its details 
below. 

Once the full solution is constructed, one may study the properties of the multiple interface system 
using the dictionary of the AdS/CFT correspondence. We first note the diagonal and off-diagonal parts of the 
solutions involve specific homogeneous terms. We shall show that these homogeneous solutions are related to the coordinate
transformations that preserve the original BTZ form of the metric to the $O(\gamma^2)$. 
Next for the above mentioned solutions, we shall construct the Fefferman-Graham form of the 
metric and identify the energy momentum tensor of the boundary field theory using the holographic dictionary \cite{de Haro:2000xn}. 
In addition to the energy momentum contribution of the undeformed CFT,  we find the Casimir energy and 
 pressure coming from any pairs of two interfaces.  Using the boundary-horizon map in \cite{bgj,Bhattacharyya:2008xc}, we 
compute the horizon length of the multiple interface system and identify the entropy  of the interface geometry.  

The dual boundary field theory is given by the multiple-interface deformation\footnote{The 
two impurity Kondo 
system in Refs.~\cite{Affleck:1991yq,Affleck:1995ge,Malecki:2010pp} has some similarity with  our multiple interface system.} 
of the CFT on $R^{1,1}$. 
Each 
interface is realized as a jump of
the coupling  where the corresponding marginal operator is the Lagrange density operator. 
Using the conformal perturbation theory,
we shall compute  the free energy up to the $O(\gamma^2)$ in the strong coupling limit. Once the free energy is obtained,
the energy and the entropy follow from the thermodynamic relations. We shall find the agreement of these two 
independent computations in the strong coupling limit. We shall also  comment on the properties of interface lattice 
\cite{Kachru:2009xf,Horowitz:2012ky,Iizuka:2012dk,Horowitz:2012gs}
and 
discuss the zero temperature limit of the multiple interface system.

In Section 2, we  review the Janus and black Janus solutions in three dimensions, which are relevant for 
the subsequent discussions. We also present the global Janus solution which is dual to the interface deformation 
of the CFT on $S^1 \times R $. Using the method of the holographic entanglement entropy, we identify the entropy 
 of the interface on the circle $S^1$. Section 3 discusses the perturbative setup for the general static solutions 
around the BTZ black hole background.  In Section 4, we discuss the leading scalar part of solutions for the general interface system. 
Section 5 deals with the construction of the geometric part of solution to the $O(\gamma^2)$. Section 6 presents the (finite temperature)
physics of the double interfaces. The field theoretic treatment of the double interface system is considered in Section 7.
Section 8 deals with the properties of the general multiple interface systems including the interface lattice. 
In Section 9, we take the zero temperature limit of the
solutions. Last section is devoted to the discussions and  concluding remarks. Various technical details are collected in Appendices.

\section{Janus system and some known solutions}\label{sec2}
We begin with the 3d Einstein scalar system
\eq
S={1\over 16\pi G}\int d^3 x \sqrt{g}\left( R- g^{ab}\partial_a \phi \partial_b \phi + 2\right),
\eqx
which can be consistently  embedded into the type IIB supergravity theory \cite{Bak:2007jm}. Here we 
set the AdS radius $R_{AdS_3}$ to be unity  for simplicity and recover it whenever it is necessary.
We are interested in the most general geometries  for which asymptotic values of the scalar field are allowed
to vary. 
The Einstein equation  reads
\eq\label{eqofma}
R_{ab}+2 g_{ab} = \partial_a \phi \partial_b \phi 
\eqx
and the scalar equation of motion  is given by
\eq\label{eqofmb}
\partial_a (\sqrt{g} g^{ab} \partial_b \phi)=0 .
\eqx
Any resulting solutions involving nontrivial scalar field will 
be  deformations of the well known AdS$_3 \times S^3$  background \cite{de Boer:1998ip}.

\subsection{Janus solutions}\label{sec2-1}
For the static Janus solution,  we take an ansatz for the three dimensional metric  and the scalar field in the following form
\eq\label{threedmet}
ds^2= dr^2 + f(r)  {-dt^2 + d{\xi}^2\over \xi^2}, \quad \quad \phi=\phi(r) .
\eqx
Using this ansatz, 
it was found that the solution of the equations of motion (\ref{eqofma}) and (\ref{eqofmb})
is  given by  \cite{Bak:2007jm}
\eq\label{fforma}
f(r) = {1\over 2}\Big( 1+ \sqrt{1-2 \gamma^2} \cosh 2r \Big)
\eqx
and
\eq
\phi(r) = \phi_0 + 
{1\over \sqrt{2}} \log\left( {1+ \sqrt{1-2 \gamma^2} +\sqrt{2}\gamma \tanh r  \over  1+ \sqrt{1-2 \gamma^2} -\sqrt{2}\gamma \tanh r }\right) .
\eqx
Without loss of generality, we may set $\phi_0=0$.
The Janus solution holographically realizes an ICFT where two CFTs defined on $1+1$ dimensional half spaces are 
glued together over a $0+1$ dimensional interface.
The conformal boundary of the metric (\ref{threedmet}) has three components.  As $r\to \pm \infty$ and finite $\xi>0$, 
we can strip off the $1/\xi^2$ factor, and the corresponding  boundary components are 
two copies of $ R_+ \times R$ spanned by $\xi, t$.
Note that the scalar approaches two constant values at these boundaries
\eq
\label{e.phias}
\lim_{r\to \pm \infty} \phi(r) = \pm {1\over \sqrt{2}} 
\log\left( {1+ \sqrt{1-2 \gamma^2} +\sqrt{2}\gamma \over  1+ \sqrt{1-2 \gamma^2} - \sqrt{2}\gamma }\right) .
\eqx
The value of the scalar  on the boundary is dual to a coupling (modulus) of the two dimensional CFT. The third 
boundary component is at $\xi=0$, which is the boundary of the AdS$_2$ factor. This describes  the interface where the two half planes are 
glued together. Hence the dual CFT is an ICFT where two CFTs defined on a half line are at different points in their coupling space.

After introducing an angular coordinate $\mu$ by
\eq
dr= \sqrt{f(r)}d\mu ,
\eqx
we can present the above solution in the form \cite{Bak:2007jm}
\eq\label{threedmetone}
ds_3^2= f(\mu)\left(d\mu^2+  {-dt^2 + d{\xi}^2\over \xi^2}\right), \quad \quad \phi=\phi(\mu)
\eqx
where
\eqn
\label{efmuone1}
f(\mu)&=&\f{\kappa_+^2}{{\rm sn}^2 (\kappa_+ (\mu+\mu_0),k^2)}=\f{\kappa_+^2{\rm dn}^2 (\kappa_+ \mu,k^2) }{{\rm cn}^2 (\kappa_+ \mu,k^2)} , 
\\
\phi(\mu)&=& \f{1}{\sqrt{2}}\ln \left( \f{{\rm dn}(\kappa_+ (\mu+\mu_0),k^2)-k{\rm cn}
(\kappa_+ (\mu+\mu_0),k^2)}
{{\rm dn}(\kappa_+ (\mu+\mu_0),k^2)+k{\rm cn}
(\kappa_+ (\mu+\mu_0),k^2)}
\right)
\nonumber\\
&=&\f{1}{\sqrt{2}}\ln\left(\f{1+ k\, {\rm sn} (\kappa_+ \mu,k^2)}{ 1- k\, {\rm sn} (\kappa_+ \mu,k^2)}\right)
\label{efmuone2}
\eqnx
with
\eqn
&&\kappa^2_\pm \equiv \f{1}{2} (1\pm\sqrt{1-2\gamma^2}) ,\\
&& k^2 \equiv \kappa^2_-/\kappa^2_+={\gamma^2\over 2}+O(\gamma^4) ,\\
&& \mu_0 \equiv K(k^2)/\kappa_+ =\f{\pi}{2} \left(1+ \f{3}{8}\gamma^2+ O(\gamma^4)\right) .
\eqnx
This describes Janus deformation of the Poincare patch geometry.

For the deformation of
the global AdS, we need to replace the AdS$_2$ part in (\ref{threedmet}) and (\ref{threedmetone})
by the global AdS$_2$  leading to the solution
\eq
\label{threedmettwo}
ds_3^2= f(\mu)\left(d\mu^2+  { d{\lambda}^2  -dt^2 \over \cos^2\lambda}\right),
\quad \quad \phi=\phi(\mu) ,
\eqx
where the coordinate $\lambda$ is ranged over $[-\pi/2, \pi/2]$.
The constant time slice takes a shape of a disk whose boundary is consisting of two half circles
with $\mu=\pm \mu_0$. Hence the boundary spacetime  corresponds to $S^1 \times R$ on which the ICFT
is defined. We  parameterize the circle $S^1$ by
angle $\theta$ defined by
\eqn
\theta = \left[
\begin{array}{cl}
\pi/2-\lambda & {\rm for} \  \mu=\mu_0 , \\
\lambda-\pi/2 & {\rm for} \  \mu=-\mu_0 .
\end{array}\right.
\eqnx
For the undeformed case  with $\phi=0$, the metric takes a form
\eq
\label{threedmetzero}
ds_3^2= {1\over \cos^2 \mu}\left(d\mu^2+  { d{\lambda}^2  -dt^2 \over \cos^2\lambda}\right)
\eqx
where $\mu \in [-\pi/2, \pi/2]$. After taking the coordinate transformation,
\eq
\cosh \rho ={1\over \cos\mu\cos\lambda}\,, \ \ \ \cos \theta= \sin \lambda\, \coth \rho\,,
\eqx
one is led to the usual form of the metric
\eq
ds^2= -\cosh^2\rho dt^2 + d\rho^2 + \sinh^2\rho d\theta^2\,,
\eqx
where $\theta \in [-\pi, \pi]$.

The global Janus background is dual to the interface CFT on  $S^1 \times R$.
Using the standard holographic dictionary, the energy density and the pressure
of the ICFT can be identified as
\eq
{\cal E}=p= -\f{c}{12} \,\f{1}{ 2\pi}
 ,
\eqx
where we have used the general fact
\eq
\frac{R_{AdS_3}}{4G} =\frac{c}{6} .
\eqx 
Note that these quantities are independent of the Janus deformation parameter and agree with the well known results for the CFT
on $S^1 \times R$ where the circle size is set to be $2\pi$.

\subsection{Entanglement entropy of global Janus}
A useful observable in the ICFT is the entanglement entropy which is defined as follows.  
The space  on which the CFT is living is divided into two  regions ${\cal A}$ and ${\bar{\cal A}}$. 
The total Hilbert space of states $H$ is expressed by the product $H=H_{\cal A}\otimes H_{\bar{\cal A}}$, 
where $H_{{\cal A}}$ ($ H_{\bar{\cal A}}$) is supported on ${\cal A}$  (${\bar{\cal A}}$).
 A reduced density matrix can be defined by tracing over all states in ${\bar{\cal A}}$,
\eq
\rho_{\cal A} = {\rm tr}_{{\bar{\cal A}}} \rho\, ,
\eqx
where $\rho$ is the density matrix of the total system. At zero temperature, one takes $\rho$ to be  
the projector on the ground state. The entanglement entropy  associated with the region ${\cal A}$ is then defined as
\eq\label{entangent}
S_{\cal A} =- {\rm tr}_{{\cal A}} \rho_{\cal A} \log \rho_{\cal A}\, .
\eqx

A holographic prescription to calculate the entanglement entropy in  spaces which are asymptotic to AdS$_{d+1}$ was presented  in \cite{Ryu:2006bv,Ryu:2006ef}.
 We denote the boundary of the region ${\cal A}$ by $\partial {\cal A}$.
A static minimal surface ${\Sigma}_{\cal A}$ extends into 
the AdS$_{d+1}$ bulk and ends on $\partial {\cal A}$
as one approaches the boundary of AdS$_{d+1}$.
The holographic entanglement entropy can then be calculated as
\eq
S_{\cal A} = {{\rm Area}({ \Sigma}_{\cal A}) \over 4 G_{d+1}} ,
\eqx
where ${\rm Area}({\Sigma}_{\cal A})$ denotes the area of the minimal surface ${\Sigma}_{\cal A}$ and $ G_{d+1}$ is
the Newton constant for AdS$_{d+1}$ gravity.

The minimal surface is a space-like geodesic connecting the points in the constant-time slicing
given by
\eq
dr^2 + {f(r)\over \cos^2\lambda} d\lambda^2\,.
\eqx
The geodesic to compute the entanglement  entropy chooses
the $\lambda$ coordinate as  constant $ \lambda =\pi/2-\theta_0$,
while $r$ varies from $- \infty$ to $+ \infty$. This corresponds to a symmetric region  of $-\theta_{0}\le \theta\le \theta_0$
around an interface at $\theta=0$.

The geodesic length is divergent and has to be regularized by introducing a cutoff $\delta$  near the boundary \cite{Azeyanagi:2007qj}
 \eq
\label{geolength}
{\rm Area}(\Sigma) = R_{AdS_3}\int^{r_{\infty}}_{r_{-\infty}} dr=
R_{AdS_3}\big(r_{\infty} (\Sigma) - r_{-\infty} (\Sigma)\big) .
\eqx
The regularized length can be read off from the relation
$f(r_{\pm \infty})/\cos^2 \lambda =f(r_{\pm \infty})/\sin^2 \theta_0 \approx 1/\delta^2$
which leads to
\eq
r_{\pm \infty} = \mp\Big(\log \delta  +{1\over 2} \log  \sqrt{1-2\gamma^2}- \log(2\sin\theta_0)\, \Big) .
\eqx
Hence
\begin{eqnarray}\label{gfachol}
{\rm Area}(\Sigma) &=&R_{AdS_3}\Big( r_{\infty} (\Sigma) - r_{-\infty} (\Sigma) \Big) \nonumber\\
&=&R_{AdS_3} \Big(  2 \log {2\sin\theta_0 \over \delta} -\log \sqrt{1-2\gamma^2}\Big) .
\end{eqnarray}
The holographic result 
\eq
S_{\cal A} = {c\over 3} \log {2 \sin\theta_0 \over \delta} + \log g_{I}
\label{gentropy}
\eqx
has the same general form as the entanglement entropy calculated on the CFT
 side using the replica trick \cite{Calabrese:2004eu},
where  $\delta$ is  the UV cutoff and 
\eq
S_I = \log g_I =\f{c}{6} \log \f{1}{\sqrt{1-2\gamma^2}}\,.
\eqx
This  is interface entropy (sometimes called  g-factor \cite{Affleck:1991tk})  
which is associated with the degrees of freedom localized on the interface.  Our 2d bulk part, i.e. $(S_{\cal A}- S_I)$, agrees with the general  CFT result in \cite{Calabrese:2004eu}.
In \cite{Azeyanagi:2007qj}, using the replica trick of CFT, the interface entropy in the weak coupling 
limit has been computed as
\eq
S'_I = \log g'_I =\f{c}{48} \log \f{1}{2}\left[\left(\f{1+\sqrt{2}\gamma}
{1-\sqrt{2}\gamma}\right)^{\f{1}{2\sqrt{2}}}+
\left(\f{1-\sqrt{2}\gamma}{1+\sqrt{2}\gamma}\right)^{\f{1}{2\sqrt{2}}}\right].
\eqx
This agrees with the above gravity result, which is inherently  in the strongly coupled limit, 
only up to $O(\gamma^2)$. Since the computation in the strongly coupled side is compared with
that in the weakly coupled side, there is no reason to expect any agreement of the two.
Below in the field theory computation, we shall use the conformal perturbation theory in which the basic information about
the coupling is contained in the correlation functions of relevant operators. Since we use the forms of correlation functions 
in the strong coupling  limit, it is expected to have an agreement of the two sides even at higher orders of $\gamma$. However,
our check below will be up to $O(\gamma^2)$.
\section{Black Janus as a perturbation}\label{sec3}

We  begin our discussion of three dimensional Janus black holes by studying the leading
order corrections to the  geometry and the scalar field starting from the  BTZ black
hole solution.
The Euclidean BTZ black hole  in three dimensions  \cite{Banados:1992wn} can be written as
\eq
ds^2= \f{1}{z^2} \left[ (1-z^2) d\tau^2+dx^2+\f{dz^2}{1-z^2} \right] 
\label{btz}
\eqx
where the coordinate $x$  is range over  $(-\infty,\,\infty)$.
Of course the $x$ direction can be compactified on a circle  but we shall be concerned
below only with the non compact case.  Note that the horizon is located at $z=1$.   
The regularity near $z=1$ is ensured if the Euclidean time coordinate $\tau$ has a period $\beta=2\pi$.
The corresponding Gibbons-Hawking temperature can be identified as
\eq
T={1\over 2\pi} .
\eqx
The BTZ black hole with a general temperature can be given
by the metric
\eq
ds^2= \f{1}{{z'}^2} \left[ (1-a^2\,{z'}^2)
d{\tau'}^2+d{x'}^2+\f{d{z'}^2}{1-a^2\,{z'}^2} \right] 
\eqx
which may be obtained by the scale coordinate transformation
\eq
z'= \frac{z}{a}, \quad \tau'=\frac{\tau}{a}, \quad   x'=\frac{x}{a}
\eqx
from (\ref{btz}).
The temperature for this scaled version  now becomes
\eq
T'= {a\over 2\pi } .
\eqx
Below we  shall  mostly take the temperature $T=(2\pi)^{-1}$ for the sake of simplicity.  
When it is necessary,  we shall recover the general temperature dependence 
using this scale coordinate transformation.

\subsection{Linearized deformation}

Introducing a new coordinate $y$ given
by $z=\sin y$, the planar black hole metric (\ref{btz}) can be rewritten as
\eq
ds^2=\f{1}{\sin^2 y} \left[\, \cos^2 y \, d\tau^2+dx^2+dy^2 \right] .
\eqx
Motivated by the form of this metric, we shall make the following ansatz
\eq
ds^2=  {dx^2 + dy^2\over A(x,y)}+{d\tau^2\over B(x,y)},
\quad \quad \quad \phi=\phi(x,y) ,
\eqx
which describes general static geometries. 
It is then straightforward to show that the equations of motion (\ref{eqofma}) and (\ref{eqofmb})  reduce to
\eqn
&&({\vec\partial} A)^2 -A\, {\vec\partial}^2 A=2A - A^2 \,({\vec\partial} \phi)^2 ,\\
&& 3({\vec\partial} B)^2 -2 B\, {\vec\partial}^2 B=8 B^2/A ,\\
&& {\vec\partial} B \cdot {\vec\partial} \phi -2 B \, {\vec\partial}^2 \phi=0 ,
\eqnx
where we introduced the notation
${\vec\partial}=(\partial_{x},\partial_{y})$.

As a power series in $\gamma$, the scalar field may be expanded as
\eq
\phi(x,y)=\sum^\infty_{n=0}\gamma^{2n+1} \phi_{2n+1}(x,y) .
\eqx
Then the scalar equation in the leading order becomes
\eq
\tan y \,{\partial_y} \varphi -\sin^2 y\, {\vec\partial}^2 \varphi=0
\label{eq0}
\eqx
where $\varphi(x,y)$ denotes $\phi_1(x,y)$. In the next section, we will construct the most general solution
of this equation. 

The leading perturbation
of the metric part  begins at $O(\gamma^2)$. Let us organize the series expansions of 
the metric variables by
\eq
A=A_0 \Big(1+{\gamma^2\over 4}a(x,y)+O(\gamma^4)\Big), \quad
B=B_0 \Big(1+{\gamma^2\over 4}b(x,y)+O(\gamma^4)\Big),
\eqx
where
\eq
A_0 =\sin^2y, \quad  B_0 = \tan^2y .
\eqx
The leading order equations for the metric part then become
\eqn
&& -2 a +\sin^2y\, {\vec\partial}^2 a
=+4 \sin^2 y  ({\vec\partial} \varphi)^2,
\label{eq1}
\\
&& - 2 \tan y \,{\partial_y} b +\sin^2 y\, {\vec\partial}^2 b  =+4 a .
\label{eq2}
\eqnx
These linear partial  differential equations (with the source term), (\ref{eq0}), (\ref{eq1}) and (\ref{eq2}) are of our main interest below.

\subsection{Linearized Black Janus}

Using the Janus boundary condition
$\phi(x,0)= \gamma\,  {\epsilon}(x)+O(\gamma^3)$ with the sign function $\epsilon(x)$,
the leading order scalar equation is solved by \cite{bgj}
\eq
\label{e.scalarlin}
\varphi=\f{\sinh x}{\sqrt{\sinh^2 x +\sin^2 y} } .
\eqx
The solution for the geometry part can be found as $a(x,y)=b(x,y)=q(x,y)$ where
 \eq
q(x,y)=3 \left(\f{\sinh x}{\sin y}\right) \,\tan^{-1}\left(\f{\sinh x}{\sin y}\right)
+\f{\sinh^2 x}{\sinh^2 x+\sin^2 y}+2 +2 C \,  {\sinh x \over \sin y}
\label{qsolu0}
\eqx
with a $O(1)$ integration constant $C$ \cite{bgj}.
Indeed checking that (\ref{qsolu0}) solves
Eqs. (\ref{eq1}) and (\ref{eq2}) is straightforward.
Then the metric for the black Janus can be written as
\eq
ds^2=  {1-\f{\gamma^2}{4} q(x,y)\over \sin^2 y}\left[\, \cos^2y \, d\tau^2+dx^2+dy^2 \right]+O(\gamma^4)
.
\label{bmetric0}
\eqx
Next we  introduce  an  angular coordinate $\mu$ that is defined by
\eq
\tan \Big(\mu+ {\gamma^2\over 4} C \Big) = \f{\sinh x}{\sin y} .
\eqx
The above metric for the linearized black Janus can be written using the scale function $f(\mu)$
of the original Janus solution: Namely,
 the metric can be expressed in the following form
\eqn
 ds^2=\f{f\big(\mu\big)}{\sinh^2 x+\sin^2 y} \left[\, \cos^2 y \, d\tau^2+dx^2+dy^2 \right]+O(\gamma^4) 
\label{bmetric}
\eqnx
with $f(\mu)$  given in (\ref{efmuone1}).
To show this, we have used the expansion of the scale function $f(\mu)$  in the form
\eq
f\big(\mu\big)={1-\f{\gamma^2}{4} q(x,y)\over \cos^2 \Big(\mu+ {\gamma^2\over 4} C \Big)}
+O(\gamma^4) .
\eqx
The zeroes of the function $A$ and $B$ occur at $\mu=\pm \mu_0$,
which correspond to the boundary of the asymptotically AdS space. As a consequence the coordinate $\mu$ is ranging
over $[-\mu_0,\, \mu_0]$ as in the case of the original Janus solution.

In fact the all order exact black hole solution with the Janus boundary condition can be found as \cite{bgj}
\eq
\label{threemetthree}
ds^2= f(\mu)\left(d\mu^2+  { d\kappa^2  -dt^2 \sinh^2 \kappa}\right),
\quad \quad \phi=\phi(\mu)
\eqx
where the coordinate $\kappa$ is ranged over $[0, \infty)$.
The horizon is located at $\kappa=0$  whereas $\mu=\pm \mu_0$, and  $(\mu=0,\, \kappa=\infty)$ correspond
to the boundary  on which the ICFT is defined.

\section{Scalar part for  multiple interfaces}

In this section, we shall analyze the linearized equation (\ref{eq0}) 
for the general cases. We begin with the scalar field perturbation. The equation can be rewritten 
as
\eq
\Big[ \partial_x^2 + 4 s\partial_s\, (1-s)\, \partial_s \Big] \varphi(x,s)=0
\eqx
where we introduce the variable $s$ by $s=\sin^2 y$.
This is solved by
\eq
\varphi (x,s)= \int d k\,\tilde{\varphi}(k)\, e^{ikx}  F(ik/2, -ik/2; 1; 1-s) |\Gamma(1+ik/2)|^2
\label{fintegral}
\eqx
where $F(a,b;c;x)$ is the hypergeometric function
\eq
F(a,b;c;z)=\sum^\infty_{n=0} \f{(a)_n (b)_n}{(c)_n} \frac{z^n}{n!}
\label{hyperseries}
\eqx
with the Pochhammer symbol $(a)_n=a(a+1)(a+2)\cdots (a+n-1)$ ($(a)_0=1$). Note
\eq
F(a,b;c;1)={\Gamma(c)\Gamma(c-a-b)\over \Gamma(c-a)\Gamma(c-b)}
\eqx
if ${\rm Re}(c)> {\rm Re}(a+b)$ and thus
\eq
F(ik/2, -ik/2; 1; 1) |\Gamma(1+ik/2)|^2 =1 .
\eqx

Let us first look at the behavior of the scalar perturbation at the boundary $s=0$. 
Since the scalar field at the boundary has to be well behaved and finite, it can be 
expanded as
\eq
\varphi = \varphi_0 (x) + s \varphi_1 (x)+s^2  \varphi_2 (x) +\cdots .
\eqx
Then the scalar equation tells us that
\eq
\partial_x^2 \varphi_0 (x)  + s \Big[
(\partial_x^2 -2) \varphi_2(x) + 8 \varphi_4(x) 
\Big]+ O(s^2)=0 .
\eqx
Therefore $\varphi_0 (x)$ has to be piecewise constant since the
term linear in $x$ violates the requirement of  asymptotically AdS space. We claim that
any piecewise constant boundary condition leads to well defined asymptotically AdS
geometry. But the rigorous  proof of this from the geometric consideration is far 
from obvious and beyond the scope of
this paper. Based on the AdS/CFT
correspondence, the corresponding regular geometry should exist since the finite 
temperature version of the field theory
with such deformation by any marginal operator is 
well defined.    

Let us illustrate various cases of solutions.
First we consider the Janus boundary condition $\varphi(x,0)=\epsilon(x)$.
For this case, one finds $\tilde{\varphi}(k)= 1/(\pi i k)$ leading to
\eq
\varphi (x,s)= \int^\infty_0 d k {2 \sin kx \over \pi k}   F(ik/2, -ik/2; 1; 1-s) |\Gamma(1+ik/2)|^2 .
\label{jscalar}
\eqx
Noting $F(ik/2, -ik/2; 1; 0)=1$ and  $|\Gamma(1+ik/2)|^2 ={\pi k/2 \over \sinh {\pi k/2}}$, 
one can perform the integral for $s=1$ and find 
\eq
\varphi (x,s=1)= \int^\infty_0 d k {\sin kx \over \sinh \pi k/2}   =\tanh x .
\label{singlej}
\eqx
Then using this and the series expansion of the hypergeometric function in 
(\ref{hyperseries}), the expression in (\ref{jscalar}) can be arranged and summed up as
\eq
\varphi (x,s) =
\sum^\infty_{n=0}\frac{(1-s)^n}{(n!)^2}\Big({-}\frac{\partial_x}{2}\Big)_n\Big(\frac{\partial_x}{2}\Big)_n
\tanh x =
{\sinh x \over \sqrt{\sinh^2 x +s}}
\eqx
which is nothing but (\ref{e.scalarlin}). 

The second is an example where the boundary condition  $\varphi(x,0)$ is given by
\eq
 \varphi(x,0)=\left[
\begin{array}{ll}
2, &  {\rm for}\    0 \le x  \le l ,\\
0,  & {\rm otherwise}.
\end{array}
\right.
\label{two-defect}
\eqx
The corresponding scalar solution can be found as
\eq
\varphi (x,s)= 2 \int^\infty_0 d k { \cos k(x-l/2)\sin kl/2 \over \sinh \pi k/2}   F(ik/2, -ik/2; 1; 1-s)
\eqx
leading to
\eq
\varphi (x,s=1)= {2\sinh l \over \cosh (2x-l) + \cosh l}= \tanh x -\tanh (x-l) .
\eqx
This is a double interface system which we shall consider in detail later on.

\subsection{Realization of holographic lattice}
One can also consider the periodic case with a period $2l$. For simplicity we take the simple case
\eq
 \varphi_L(x,0)=\left[
\begin{array}{cl}
1, &  {\rm for}\ \  \  0 \le x   < l ,\\
-1,  & {\rm for}\  -l  \le x  < 0 .
\end{array}
\right.
\label{latticebc}
\eqx
With $k_m = \pi (2m+1)/l$, the scalar solution can be expressed as a Fourier series
\eq
\varphi_L (x,s)= {2\pi\over l}\sum^{\infty}_{m=0}  {\sin k_m x \over  \pi k_m/2} 
  F(ik_m/2, -ik_m/2; 1; 1-s)  |\Gamma(1+ik_m/2)|^2 . \label{fsum}
\eqx
On the horizon at $s=1$, the scalar takes the form
\eq
\varphi_L (x,s=1)= {2\pi\over l}\sum^{\infty}_{m=0}  {\sin k_m x \over  \sinh \pi k_m/2}
= \omega \omega_+  {\rm sn} \Big(\omega_+ x,\omega^2\Big) \label{jacobisn}
\eqx
where $\pi \omega_+ ={2 K'(\omega^2) } ={2 K(1-\omega^2)} $ with the complete elliptic integral $K$, and $\omega$ ($\in (0,1)$) is related to 
the periodicity by
\eq
2l = 2\pi  K(\omega^2)/K'(\omega^2)=2 \pi K(\omega^2)/K(1-\omega^2).
\eqx
As $\omega \rightarrow 1$, $l \rightarrow \infty$  the boundary condition (\ref{latticebc}) approaches that
of the black Janus. The corresponding horizon image has the limit
\eq
\varphi_L (x,s=1) \rightarrow \tanh x
\eqx
which agrees with (\ref{singlej}).
For general $s$, one finds an alternative form
\eq
\varphi_L (x,s) =\omega \omega_+
\sum^\infty_{n=0}\frac{(1-s)^n}{(n!)^2}\Big({-}\frac{\partial_x}{2}\Big)_n\Big(\frac{\partial_x}{2}\Big)_n
 {\rm sn} (\omega_+ x,\omega^2)
\eqx
with the Jacobi elliptic function $\mathrm{sn}(\omega_+ x,\omega^2)$, and it is  not simple to sum up.

\subsection{Simple form of general solutions}
\begin{figure}[t]
\centering
\includegraphics[scale=1.1]{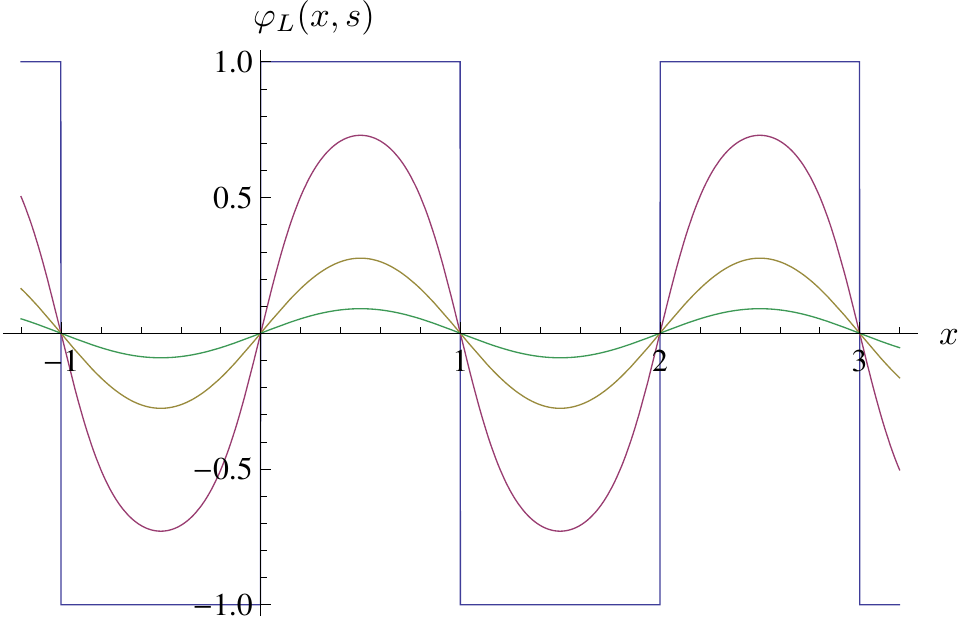}
\caption{Some plots of the function (\ref{perphi}) with $l=1$  for $s=0,\frac{1}{10},\frac{1}{2},1$ (from the above).}
\label{fig_phi}
\end{figure}

For a further analysis, it is convenient to use another alternative expression for $\varphi(x,s)$ instead
 of the Fourier integral form (\ref{fintegral}). Note that there is a translational invariance
 in the $x$ direction in (\ref{eq0}) and thus $\varphi(x-l,s)$ is also a solution when $\varphi(x,s)$ is.
The most general  solution takes the form
\eq
\varphi(x,s)=\sum_{n=-\infty}^{\infty} \alpha_n \varphi_0(x-l_n,s) \label{perphi0}
\label{gscalar}
\eqx
where  $\varphi_0(x,s)=\frac{\sinh x }{\sqrt{\sinh^2 x +s}}$ in (\ref{e.scalarlin}) and  $\alpha_n,\ l_n$ 
are constants realizing general piecewise constant boundary condition at infinity ($s=0$). We shall order $l_n$ such that
$l_n < l_{n+1}$ for all $n$ and refer $\alpha_n$ as the $n$-th interface coefficient.
For the above form of general solution, $\varphi(x,0)$  can be identified as
\eq
\varphi(x,0)=\sum_{n=-\infty}^{\infty} \alpha_n \epsilon(x-l_n) .
\label{generalbc}
\eqx
which is the most general boundary condition consistent with our previous analysis.
For  the double interfaces in (\ref{two-defect}), one is led to the rather simple expression
\eq
\varphi(x,s)= \varphi_0(x,s) -\varphi_0(x-l,s)
\label{twodil}
\eqx
We may present  the periodic solution (\ref{fsum}) in the form
\eqn
\varphi_{L}(x,s)=\sum_{n=-\infty}^{\infty} (-1)^n \varphi_0(x-nl,s) \label{perphi}
\eqnx
which is more convenient to deal with.
 In Fig.~\ref{fig_phi}, some plots of the function (\ref{perphi}) for various values of $s$ are given where  the
length $l$ is set to be unity.
Note that $\varphi(x,0)$ shows the form of  the square wave function,  which is set by the boundary condition. When $s=1$, it becomes
\eqn
\varphi_L(x,s=1)=\sum_{n=-\infty}^{\infty} (-1)^n \tanh(x-nl), \label{phi_s1}
\eqnx
which is nothing but a different representation of the Jacobi elliptic function in (\ref{jacobisn}).

\section{Geometric part}

For the geometry part, one finds that
$a$ and $b$ behave as
\eqn
&& a(x,y)=a_{H0}(x)+ \cos^2y \, a_{H1} (x)+ \cos^4y \, a_{H2}(x)+\cdots , \nonumber\\ 
&& b(x,y)=b_{H0}(x)+ \cos^2y\, b_{H1} (x)+ \cos^4y\,  b_{H2}(x)+\cdots 
\label{expansionh}
\eqnx
near horizon region, where $y=\f{\pi}{2}$ still corresponds to the horizon location\footnote{
The horizon here is defined by the surface of $g_{\tau\tau}=0$. This value $y=\f{\pi}{2}$
can be regarded as a coordinate choice since we have the freedom of coordinate transformation.
We demand the regularity of geometry on the horizon and the Dirichlet boundary condition
$\f{\partial }{\partial y} a|_{y=\f{\pi}{2}}=\f{\partial }{\partial y} a|_{y=\f{\pi}{2}}=0$ \cite{bgj}.
Then the expansions in (\ref{expansionh}) follows from a straightforward analysis of the 
differential equations (\ref{eq1}) and (\ref{eq2}).
}. 
On the other hand, near the boundary at infinity, they should have the expansion
\eqn
&&a(x,y)=\sin^2y \, a_1(x)+\sin^4y \, a_2(x)+\sin^6y \, a_3 (x)+\cdots , \nonumber\\
&& b(x,y)=b_0(x)+
\sin^2y \, b_1(x)+\sin^4y \, b_2(x)+\cdots ,
\eqnx
where
$a_0$ term is not allowed, 
and $a_1$ and $b_0$
are related to the extra mass density of black hole induced by the multiple interfaces. 
The term of order $s^{-1/2}=1/\sin y$ can be added
representing the freedom of coordinate transformations. 
We shall get back to this issue in the analysis
of examples below.

The back reaction to the gravity sector is determined by two equations, (\ref{eq1}) and (\ref{eq2}). We plug the solution 
(\ref{gscalar}) into (\ref{eq1}), getting
\eqn
4 \sin^2y (\vec\partial \varphi)^2= 4 \sin^2y 
\sum_{n_1, n_2} 
\alpha_{n_1} \alpha_{n_2}\vec\partial \varphi_0(x-l_{n_1},y)\cdot\vec\partial \varphi_0(x-l_{n_2},y)
\eqnx
for the right hand side. It is convenient to split this into the ``diagonal'' part
\eqn
4  \sin^2y\sum_{n=-\infty}^{\infty} \alpha^2_n (\vec\partial \varphi_0 (x-l_n,y))^2 \label{src1}
\eqnx
and the ``off-diagonal'' part
\eqn
4  \sin^2y \sum_{n_1 < n_2} 2 \alpha_{n_1}\alpha_{n_2}\vec\partial \varphi_0 (x-l_{n_1},y)\cdot \vec\partial \varphi_0 (x-l_{n_2},y) .
\label{src2}
\eqnx
The linearity of (\ref{eq1}) allows us to find its solution  by adding the solution with the source term (\ref{src1}) and the one with (\ref{src2}).
Each source term for the diagonal part is the same as that for the black Janus up to the translation in $x$.  
One may then easily find the solution for  the diagonal part
\eqn
a^{\rm diag}(x,y)=b^{\rm diag}(x,y)=\sum_{n=-\infty}^\infty \alpha_n^2  \, \Big[ 
q_0(x-l_n,y) + C_n \f{\sinh (x-l_n)}{\sin y}\Big] , 
\label{gdiag}
\eqnx
where
\eqn
q_0(x,y)= 3  \, \f{\sinh x}{ \sin y} \,\left(\tan^{-1}\left(\frac{\sinh x}{ \sin y}\right)-\frac{\pi}{2}\right)+2
+\f{\sin^2 y}{\sinh^2 x+ \sin^2y} .
\label{diaga}
\eqnx
 Later we shall show that the different choices of $C_n$ are all related by an appropriate coordinate transformation. 
Hence we can set $C_n=0$ without loss of generality. 

\subsection{Off-diagonal part}

Finding a closed form of the off-diagonal part is rather involved. First step in that direction should be finding
a solution of the equation with a term in the off-diagonal source term (\ref{src2}). Since 
one has a translational invariance in $x$ direction and $l>0$ is arbitrary, it suffices to consider the case of
$n_1=0$ and $n_2= 1$ with $l_0=0$ and $l_1=l$. This leads to the equations\footnote{
Here and below, the subscript or the superscript $c$ in $a_c$ and $b_c$ represents the cross-term solution of 
 (\ref{crosseqb1}) and  (\ref{crosseqb}), whose source term is from the cross-term contribution of the scalar field of the two interfaces 
that  involve unit interface strengths and are located respectively at $x=0$ and $x=l$. 
}
\eqn
-2a_c(x,y,l)+ \sin^2y\,\, \vec\partial^2 a_c(x,y,l) &=&  \frac{4(\cosh l+XY)}{(1+X^2)^{3/2}(1+Y^2)^{3/2}},
\label{crosseqb1}
\\
-2\tan y\,\, \partial_y\, b_c(x,y,l) +  \sin^2y\,\, \vec\partial^2 b_c(x,y,l) &=& 4a_c(x,y;l),
\label{crosseqb}
\eqnx
where we have introduced 
\eq
X=\f{\sinh x}{\sin y},  \quad  Y=\f{\sinh(x-l)}{\sin y}.
\eqx
Once the solution of the above equations is given, 
the full off-diaginal part of the solution may be given in the form
\eqn
&& a^{\rm off}(x,y)= \sum_{n_1 < n_2} 2 \alpha_{n_1}\alpha_{n_2} \, a_c(x-l_{n_1},y, l_{n_2}-l_{n_1}),\nonumber\\
&& b^{\rm off}(x,y)= \sum_{n_1 < n_2} 2 \alpha_{n_1}\alpha_{n_2} \, b_c(x-l_{n_1},y, l_{n_2}-l_{n_1}) .
\label{offdiag}
\eqnx

We first take the range of $x$ and $y$
as $ l< x$ and $0\le \sin y \le 1$. Then $X$ and $Y$ can take positive real values. 
We shall describe below how to extend the solution of this region
to  $ x \le l$. 

Remarkably, it turns out that we can find a solution of this equation by taking the ansatz\footnote{
One may motivate this ansatz by studying the scaling behavior of the solution near $x=y=0$ and $x-l=y=0$.  We do not 
know why the problem  is reduced to an ordinary differential equation with just one variable $m$.}:
\eqn
a_c(x,y,l)= \frac{1}{\sqrt{1+X^2}\sqrt{1+Y^2}} G(m) , \quad
b_c(x,y,l)=H(m)\label{ansatz1}
\eqnx
where
\eqn
m=(\sqrt{1+X^2}-X)(\sqrt{1+Y^2}-Y) .
\eqnx
Using this ansatz and regarding $m$ and $XY$ as two independent variables, we may cast the above equation into the form:
\eqn
\Gamma_2 \, G''(m) +\Gamma_1 \, G'(m) +\Gamma_0\, G(m)= 4(\cosh l +XY) \label{eqF}
\eqnx
where 
\eqn
\Gamma_2&=& \frac{1}{2}(1+m^2)^2 +m(1+m^2)\cosh l + m(1+m^2 + 2m\cosh l )\,XY  ,\\
\Gamma_1&=&\frac{1}{4}{(3+m^2)(m+m^{-1})} +2 \cosh l +\frac{1}{2}(1+3m^2+4m\cosh l )\, X Y ,\\
\Gamma_0&=&-\frac{1}{4}{(1+3m^2)(1+3m^{-2})}-(3m+3m^{-1}+2\cosh l )\, XY .
\eqnx
Since we treat $XY$ and $m$ independent, (\ref{eqF}) implies two ordinary differential equations:
\eqn
4&=& (m^3+m + 2m^2 \cosh l)G''
 +\f{1}{2} \Big( {3m^2+1} -4m \cosh l \Big)G'\nonumber\\
&&-(3m +3m^{-1} + 2 \cosh l )G\,, \\
-16 \cosh l&=&\Big(2(m^2+1)^2 +4m(m^2+1)\cosh l \Big)G''+(m^3+4m+3m^{-1} +8 \cosh l)G'\nonumber\\
&& -(3m^2+10+3 m^{-2})G .
\eqnx
We may eliminate the second order derivative term and get the following first order equation:
\eqn
&&2m(1-m^2)\left(1+m^2+2m \cosh l \right) G'(m)+\left(3+2m^2+3m^4 +4m(1+m^2)\cosh l \right)G(m) \nonumber\\
&& \ = 8m(1+m^2+2m\cosh l). \label{eqF2}
\eqnx
Indeed one can check that any solution of this first order equation solves the above two second order equations at the same time.
Using the homogeneous solution to this equation,
\eqn
G_h(m,\cosh l)=\frac{1}{m^{3/2}}(1-m^2)\sqrt{1+m^2+2m\cosh l},
\eqnx
one may easily find the solution\footnote{The $\cosh l$ dependence will be omitted below in case it is not confusing.}
\eqn
G(m)&=& 4 G_h(m)\int_{0}^{m} \frac{dx}{(1-x^2)G_h(x)}  \nonumber \\
&=& 4 G_h(m)\int_{0}^{m}  dx \frac{ x^{3/2}}{(1-x^2)^2\sqrt{1+x^2+2x \cosh l}}\, .
\eqnx
Using the identity,
\eqn
{x^{3\over2}\over (1-x^2)^2 \sqrt{1+x^2+2x \cosh l}}&=&\left(
{x^{5\over2}\over 2 (1-x^2) \sqrt{1+x^2+2x \cosh l}}
\right)'\nonumber\\
&&\quad -{1\over 4}\left({x\over 1+x^2+ 2x \cosh l}\right)^{3\over2} ,
\eqnx
the solution can be presented in the form
\eq
G(m)={\cal G}(m,\cosh l)\equiv 2m - {\cal I}(m,\cosh l)G_h(m,\cosh l) , \label{solf}
\eqx
where the integral ${\cal I}(x,\cosh l)$ is given by
\eq
{\cal I}(m, \cosh l)=\int_{0}^{m} dx\left({x\over 1+x^2+ 2x \cosh l}\right)^{3\over2} .
\eqx
Of course one may add the homogeneous part of the solution $C_G G_h(m)$ to the above with $C_G$ being an integration constant.
We note that  the integral ${\cal I}(m)$  can be evaluated explicitly in terms of the elliptic functions.
The details are given in Appendix A.

\subsection{Remaining off-diagonal part}

Now we turn to the second equation, (\ref{crosseqb}) for the metric.
Again the linearity of the equation allows us to find the desired solution by summing every solution associated with each term in the right hand side of the equation. With the ansatz in (\ref{ansatz1}),
 (\ref{crosseqb}) becomes
\eqn
2m\left(1+m^2+2m \cosh l \right)H''(m)+\left(5m^2-1+4m \cosh l\right)H'(m)= 8G(m)
\label{eqH}
\eqnx
where we use the general solution 
\eq
G(m)= {\cal G}(m) +C_G \, G_h(m)
\eqx
constructed in the previous section.
This equation is essentially first order with respect to $H'$. 
We introduce the homogeneous solution $Q$ given by 
\eqn
Q(m)=\frac{m^{\f{1}{2}}}{(1+m^2+2m \cosh l )^{\f{3}{2}}},
\eqnx
Then Eq.~(\ref{eqH}) can be rewritten as
\eqn
  \left(
\frac{H'}{Q}
\right)'=4  \frac{\left(1+m^2+2m\cosh l \right)^{\f{1}{2}} }{m^{\f{3}{2}}}\, G(m) .
\eqnx
Now note that the right hand side can be rearranged as
\eqn
  \left(
\frac{H'}{Q}
\right)'
=  \left[ 2\left(\frac{1+m^2+2m\cosh l}{m}\right)^{2}
\left({\cal I}(m)-C_G\right)\right]' +6\left(\frac{1+m^2+2m\cosh l}{m}\right)^{\frac{1}{2}}.
\eqnx
Now integral of this yields
\eqn
\frac{H'}{Q}= 2\left(\frac{1+m^2+2m\cosh l}{m}\right)^{2} \,\,\left({\cal I}(m)-C_G\right)
+6  {\cal M}(m) +8 C_G\sinh^2 l 
\eqnx
where ${\cal M}(m)$ is defined by
\eq
{\cal M}(m, \cosh l)\equiv \int_{0}^{m} dx \left( \frac{1+x^2 + 2x \cosh l}{x}\right)^{1\over2}
\eqx
and we added an integration constant $8 C_G\sinh^2 l $.
Now using the identities
\eqn
&& 3{\cal M}(x) +4 \sinh^2 l\,\,\, {\cal I}(x) =\frac{2 x^{\frac{1}{2}} }{(1+x^2+ 2x \cosh l )^{\frac{1}{2}}}\, (3+ x^2 +4x \cosh l ) \\
&& \frac{(1+x^2+2x \cosh l)^{\f{1}{2}}}{x^{\frac{3}{2}}}
- \frac{4   x^{\frac{1}{2}} \sinh^2 l}{(1+x^2+2x \cosh l)^{\f{3}{2}}}
=-\left(
\frac{2x\, G_h(x)}{ 1+x^2+2x \cosh l}
\right)'
\eqnx
one is led to
\eqn
H'(m)&=&4 \left[
\frac{2m^2}{1+m^2+2m \cosh l}+
\frac{(1-m^2) \left( -{\cal I}(m) +C_G\right)}{\sqrt{m(1+m^2+2m \cosh l)}}\,\,
\right]' .
\eqnx
One more integration  provides us with
\eqn
H(m)= 
{4m\over 1+m^2+2m \cosh l } G(m)
\eqnx
and the homogeneous part of the solution 
\eq
H_h(m)= C_{H}\int_0^m dx\,\frac{x^{\frac{1}{2}}}{(1+x^2+2x \cosh l )^{\frac{3}{2}}}+D_{H}
\eqx
can be added to the above in order to obtain   a general solution.

\subsection{Other types of ansatz}
In fact  one may try various forms of  ansatz that are similar to (\ref{ansatz1}). The first alternative is
\eqn
&& a_{1c}(x,y,l)= \frac{1}{\sqrt{1+X^2}\sqrt{1+Y^2}} G_1(1/m), \quad
b_{1c}(x,y,l)=H_1(1/m) ,\label{ansatz2}
\\
&& \quad\quad 1/m=(\sqrt{1+X^2}+X)(\sqrt{1+Y^2}+Y) .
\eqnx
With this ansatz, one has  precisely  the same equations whose solution is essentially not different
from the previous one given in  (\ref{solf})\footnote{This is clear from the beginning since the ansatz is basically 
the same as before.}. Namely the general solution  solution for $G_1(1/m)$
is given by 
\eq
G_1(1/m)={\cal G}(1/m)+ C'_G G_h(1/m) .
\eqx  
Using the identities
\eqn
 {\cal G}(1/m)={\cal G}(m)+{\cal I}(\infty) G_h(m)\,, \ \ \ \  G_{h}(1/m) = -G_h(m) ,
\eqnx
one can show that
\eq
G_1(1/m)= {\cal G}(m) + \left({\cal I}(\infty) - C'_G \right) G_h(m) .
\eqx
Similar analysis can be done  for $H_1(1/m)$. Thus no new solution is generated by this ansatz.

The second alternative one may try is the form
\eqn
&& a_{2c}(x,y,l)= \frac{1}{\sqrt{1+X^2}\sqrt{1+Y^2}} G_2(\bar{m}) ,\quad
b_{2c}(x,y,l)=H_2(\bar{m}),\label{ansatz3}
\\
&& \quad\quad \bar{m}=(\sqrt{1+X^2}-X)(\sqrt{1+Y^2}+Y) .
\eqnx
This ansatz  leads to the equations which agrees with (\ref{eqF}) and (\ref{eqH}) by the replacement:
\eq
\bar{m}\rightarrow m\,, \ \ \ XY\rightarrow -XY\,, \ \ \  \cosh l \rightarrow -\cosh l\,,\ \ \
G_2, H_2 \rightarrow  -G, - H
\eqx
Using this property,  the solution can be found as
\eq
G_2(\bar{m})= 
-{\cal G}(\bar{m}, -\cosh l) \,, \ \ \  H_2(\bar{m})= 
-{4\bar{m}\over 1+\bar{m}^2-2\bar{m} \cosh l }\,\,{\cal G}(\bar{m}, -\cosh l)
\eqx
plus the corresponding homogeneous part which does not play any role in our discussion 
below.
The last alternative one may try is the form
\eqn
&& a_{3c}(x,y,l)= \frac{1}{\sqrt{1+X^2}\sqrt{1+Y^2}} G_3(1/\bar{m}) \,, \  \  \
b_{3c}(x,y,l)=H_3(1/\bar{m})\label{ansatz4}
\\
&& \quad\quad 1/\bar{m}=(\sqrt{1+X^2}+X)(\sqrt{1+Y^2}-Y) ,
\eqnx
but the corresponding solution is essentially the same as that from the second alternative ansatz. 

Thus the solution can be in general of the form
\eq
(1-\delta )G(m)+ \delta \, G_2(\bar{m}), \quad (1- \delta)H(m)+\delta \,H_2(\bar{m})
\eqx
and we shall show that $\delta $ has to be zero to satisfy the required boundary
conditions.
For this purpose, it suffices to consider the limit where $l$ goes to zero.
In this limit the cross term should be reduced to the diagonal solution in (\ref{diaga}) up to the homogeneous part of the solution.
Note that
\eq
 \frac{1}{\sqrt{1+X^2}\sqrt{1+Y^2}} {\cal G}(m, \cosh l ) \longrightarrow \   3 X \,\left(\tan^{-1} X-\frac{\pi}{2}\right)+3
-\f{X^2}{1+X^2}=q_0(x.y)
\label{limitm}
\eqx
and
\eq
 -\frac{1}{\sqrt{1+X^2}\sqrt{1+Y^2}} {\cal G}(\bar{m}, -\cosh l )\longrightarrow  \   -{2\over 1+X^2}
\eqx
in the limit $l$ goes to zero.
Comparing (\ref{limitm}) with (\ref{diaga}), we have a precise agreement with each other. Then to satisfy the required boundary
conditions in the limit $l\rightarrow 0$,  $\delta$ has to be zero since the solution  $ -{2 \over 1+X^2}$ by itself does not
satisfy the required boundary conditions. (This conclusion does not change even if one includes 
any possible homogeneous part of solutions.)

Thus the solution for the cross term including the homogeneous part becomes
\eqn
a_c(x,y,l)&=& \frac{1}{\sqrt{1+X^2}\sqrt{1+Y^2}} \, \Big({\cal G}(m,\cosh l) + C_G G_h(m,\cosh l)\Big) ,\\
b_c(x,y,l)&=&  {4m\over 1+m^2+2m \cosh l }\, \Big({\cal G}(m,\cosh l) + C_G G_h(m,\cosh l)\Big).
\label{crosssolution}
\eqnx
Finally the full off-diagonal part  can be constructed in the form (\ref{offdiag}) using the above.

\section{Double interfaces}

\begin{figure}
\centering
\includegraphics[scale=1.1]{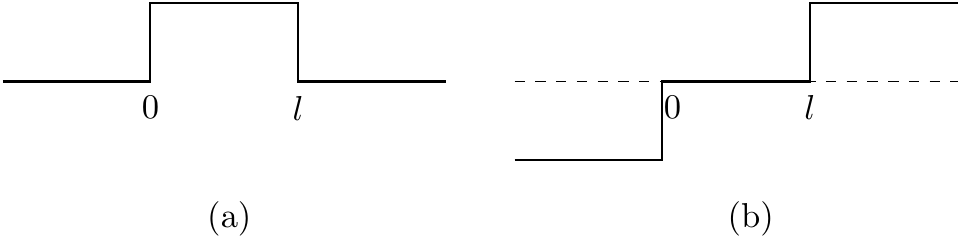}
\caption{ (a) describes the boundary condition, $\varphi(x,0)$, for $\alpha_-=-\alpha_+=1$ while (b)
for $\alpha_-=\alpha_+=1$.   }
\label{bh3}
\end{figure}

As described before, we consider the boundary condition for the scalar given by
\eq
\varphi (x,0)= \alpha_-\epsilon(x) +\alpha_+\epsilon(x-l)=\epsilon(x) -\epsilon(x-l) .
\eqx
(In Fig.~\ref{bh3}, we depict this boundary condition as well as the one 
with $\alpha_-= \alpha_+=1$.)

The scalar part of the solution has been given in (\ref{twodil}).
The full geometric part of the solution takes the form
\eqn
&&a(x,y,l)= q_0(x,y) +q_0(x-l,y) + 3\pi Y -2 a_c^0(x,y, l) , 
\\
&&b(x,y,l)= q_0(x,y) +q_0(x-l,y) + 3\pi Y -2 b_c^0(x,y, l)  ,
\eqnx
where  $a^0_c(x,y,l)$ and $ b^0_c(x,y,l)$ represents the unit-coefficient cross-term solution given by
\eqn
a^0_c(x,y,l)&=& \frac{1}{\sqrt{1+X^2}\sqrt{1+Y^2}} \,\,\,\Big({\cal G}(m,\cosh l) + {1\over 2}\Delta(l) G_h(m,\cosh l)\Big),\\
b^0_c(x,y,l)&=&  {4m\over 1+m^2+2m \cosh l }\,\, \Big({\cal G}(m,\cosh l) + {1\over 2}\Delta(l) G_h(m,\cosh l)\Big).
\label{unitcrosssolution}
\eqnx
Note that we have added here appropriate  homogeneous solutions to make the solution even under the exchange
\eq
 x \   \leftrightarrow \ l 
-x
\eqx
To see this,  we first observe that
\eq
m \leftrightarrow 1/m\,, \quad \quad X \leftrightarrow -Y
\eqx
under the exchange. 
Then one finds
\eq
{\cal G}(m)+{1\over 2}\Delta(l) G_h(m)
\eqx
is symmetric under the exchange, which one can show using the identity
\eq
{\cal G}(m)+{1\over 2}\Delta(l) \, G_h(m)=\frac{1}{2}\Big({\cal G}(m)+{\cal G}(1/m)\Big) .
\eqx
Further using the identity
\eq
q_0(x-l,y) + 3\pi Y = q_0(l-x,y) ,
\eqx
one finds that
the remaining part of the solution is also symmetric under the exchange.

\subsection{Shape of the boundary}
\begin{figure}
\centering
\includegraphics[scale=0.9]{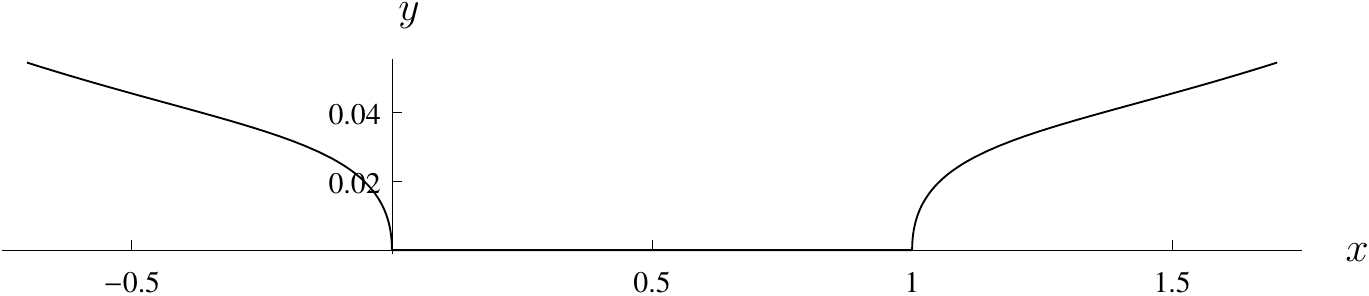}
\caption{The shape of the boundary in $(x,y)$ space is 
depicted for  $\gamma^2=0.1$, $l=1$ and  $\alpha_-=-\alpha_+=1$. The cusps 
represent the locations of interfaces. We use the convention where the coordinate $y$ grows in the upward 
direction.}
\label{boundaryfig}
\end{figure}

The change in the shape of the boundary in $(x, y)$ space is of particular interest.
Near boundary region, the diagonal and cross terms  behave completely differently.
Since the solution has the symmetry under the exchange $x \leftrightarrow l-x$, we shall describe
the behaviors of the two terms for the regions
$l \le x$ and $0 < x <l$  only. Then the remaining region $x \le 0$ will be given by the reflection symmetry.
For the region $0 < x <l$  there is no singular term in $q_0(x,y)$ and $q_0(x-l,y)+3\pi Y$. They behave
\eqn
&& q_0(x,y)=-\frac{2\sin^4 y}{5 \sinh^4 x} + O(\sin^6 y), \quad\quad {\rm for} \quad  0 \le x ,\\
&&  q_0(x-l,y) +3\pi Y= -\frac{2\sin^4 y }{5 \sinh^4 (x-l)} + O(\sin^6 y), \quad\quad {\rm for} \quad  x \le l .\\
\eqnx
On the other hand, the cross term behaves
\eqn
&& a_c(x,y) =a^c_1(x)\sin^2 y +a^c_2(x)\sin^4 y +O(\sin^6y) ,\\
&& b_c (x,y) =b^c_0(x) +b^c_1(x)\sin^2 y +O(\sin^4y)
\eqnx
for $0 < x <l$. 
Since there are no boundary terms which show singular behavior as powers of  $\sin y$, the boundary of space
 remains at $y=0$, as illustrated in Fig. \ref{boundaryfig}.

For $l \le x$, the situation is quite different: $q_0(x,y)$ term
remains nonsingular. But the terms $3\pi Y$ and  $G_h(m)$ in both of $a(x,y,l)$ and $b(x,y,l)$
involve the $1/\sin y$ singularity, which implies that  the boundary deviates from $y=0$. Let us look at some details
of  this change of boundary.  
Basically we would like to show that the singular homogeneous terms
are precisely the terms that are required for new metric (including the change of boundary) to be the form of the original
BTZ  metric without any deformation.  To show this, we begin with
\eq
ds^2=
\f{1}{\sin^2 y} \left[\frac{dx^2+dy^2}{1+\frac{\gamma^2}{4} a_h(x,y)+O(\gamma^4)} +
\frac{\cos^2 y\,\, d\tau^2}{1+\frac{\gamma^2}{4} b_h(x,y)+O(\gamma^4)}
\right] 
\eqx
where $a_h$ and $b_h$ denote the homogeneous part of the solution.
Then by the coordinate transformation,
\eqn
&& y'=y+\frac{\gamma^2}{4}\, \cos y \, \,{\cal A}_h(x,y)+O(\gamma^4) , \nonumber\\
&& x'=x+\frac{\gamma^2}{4} \cosh x \,{\cal B}_h(x,y) +O(\gamma^4) ,
\label{coortr}
\eqnx
one can bring the metric into the original BTZ form
\eq
ds^2=
\f{1}{\sin^2 y'} \left[d{x'}^2+d{y'}^2+
\cos^2 y'\, d\tau^2
\right] .
\eqx
Comparing $g'_{\tau\tau}$ and $g_{\tau\tau}$ components, one finds
\eq
{\cal A}_h(x,y)= \f{1}{2} b_h(x,y) \sin y
\eqx
together with  the conditions 
\eqn
&& \f{1}{\cosh x} \partial_x {\cal A}_h+\f{1}{\cos y} \partial_y {\cal B}_h =0,\\
&& \f{1}{\cos y} \partial_y {\cal A}_h - {\cal A}_h \sin y  - {\cal A}_h \cos^2 y =-\f{1}{2} a_h,\\
&& \f{1}{\cosh x} \partial_x {\cal B}_h + {\cal B}_h \sinh x  - {\cal A}_h \cos^2 y =-\f{1}{2} a_h.
\eqnx
Combining these equations and eliminating ${\cal A}_h$ and ${\cal B}_h$, one finds our original equations (\ref{eq1})-(\ref{eq2})
without the source term,
\eqn
&& 2a_h -\sin^2 y \,\, \vec\partial^2 a_h=0 ,\\
&& 2\tan y \,\partial_y b_h -\sin^2 y \,\,\vec\partial^2 b_h +4a_h=0 ,
\eqnx
together with the relation
\eq
a_h =b_h \sin^2 y -\sin y \cos y \partial_x b_h .
\eqx
Showing this for the homogeneous part of the cross term solution is not that straightforward
and its proof will be relegated to Appendix B. For the diagonal part, the above relation can be checked immediately
for homogeneous solution $ a^{\rm diag}_h(x,y)=b^{\rm diag}_h(x,y)= \f{\sinh x} {\sin y} $.

\begin{figure}
\centering
\includegraphics[scale=0.9]{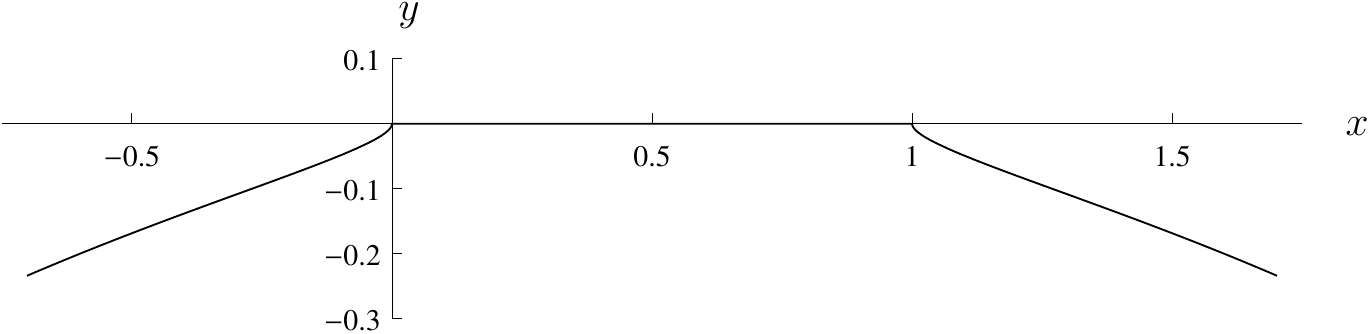}
\caption{The shape of the boundary in $(x,y)$ space is depicted for  $\gamma^2=0.1$, $l=1$ and $\alpha_-=\alpha_+=1$.
The change of the boundary is in the opposite direction compared to the configuration in Fig.~\ref{boundaryfig}.  }
\label{boundaryfigb}
\end{figure}

Thus this proves the existence of the function ${\cal A}_h$ and the corresponding coordinate transformation. The resulting form of the
coordinate transformation for $y'$  reads explicitly
\eq
y'= y +\frac{\gamma^2}{8} \left[
3\pi \sinh (x-l) \cos y  -4\Delta (l)\f{\f{1}{m}-m}{\sqrt{m+\f{1}{m}+2\cosh l}}\sin y \cos y
\right]+O(\gamma^4)
\eqx
 for $x \ge l$. 
The boundary  defined by $ y'=0$ is determined by the equation 
\eq
\sin y =\frac{\gamma^2}{8} \left[
  4\Delta (l)\f{\f{1}{m}-m}{\sqrt{m+\f{1}{m}+2\cosh l}}\sin y -3\pi \sinh (x-l) 
\right]+O(\gamma^4) .
\label{bshape}
\eqx 
For general grounds,
one expects that the curve defined by  (\ref{bshape})  can be solved in the form
\eq
\sin y = \gamma^2 g(x) + O(\gamma^4)
\eqx
with some $O(1)$ function $g(x)$.
 Consider the case where
\eqn
&& X= \frac{\sinh x}{\sin y} =  \frac{\sinh x}{g(x)} \gamma^{-2} +O(\gamma^0) \gg 1  ,\nonumber\\
&& Y= \frac{\sinh (x-l)}{\sin y} =  \frac{\sinh (x-l)}{g(x)} \gamma^{-2} +O(\gamma^0) \gg 1  .
\label{conditions}
\eqnx
 Assuming these, the curve (\ref{bshape}) is solved by 
\eq
\sin y = {\gamma^2}\left(
\Delta (l)  \sqrt{\sinh x \sinh (x-l)}-\f{3\pi}{8}\sinh(x-l) 
\right)+O(\gamma^4)
\eqx
and thus 
\eq
g(x)= \Delta (l)  \sqrt{\sinh x \sinh (x-l)}-\f{3\pi}{8}\sinh(x-l) .
\eqx
Using this, one finds that our original  conditions in (\ref{conditions}) are met if
\eq
\sqrt{x-l} \ \gg \ \gamma^2 \Delta(l) \sinh l .
\label{conda}
\eqx
For $x/l \gg 1$,  the boundary trajectory is further simplified as  
\eq
\sin y = {\gamma^2}\left(
\Delta (l) \, \, e^{\f{l}{2}}-\f{3\pi}{8}  
\right)\sinh(x-l)+O(\gamma^4) .
\eqx
For the region where the condition (\ref{conda}) is violated, one finds the solution of (\ref{bshape}) is
\eq
\sin y = O(\gamma^4) .
\eqx
Determining its precise functional form lies beyond the scope of our approximation since it requires the knowledge of  $O(\gamma^4)$ corrections to the
geometry.
The full shape of the boundary including  a solution of (\ref{bshape}) for $x > l$ is depicted in Fig.  \ref{boundaryfig}. This may
be compared with 
the boundary shape of the double interface system with 
$\alpha_-=\alpha_+=1$, which is  depicted in Fig.~\ref{boundaryfigb}

\subsection{Casimir energy and pressure}

In this subsection, we shall construct the Fefferman-Graham  metric
to determine
the boundary stress energy tensor. In order to use the prescription developed in
Ref.~\cite{de Haro:2000xn},
we introduce the metric in the
following Fefferman-Graham form,
\eq
ds^2={d\chi^2\over \chi^2}  +{1\over \chi^2} g_{\mu\nu}(q,\chi) dq^\mu dq^\nu
\label{fgform}
\eqx
where $q^\mu$ ($\mu=0,\,1$) denote the boundary coordinates
and $\chi=0$ corresponds to
the location of the boundary.
In general one may expand $g_{\mu\nu}$ by
\eq
g_{\mu\nu}(q,\chi)=g^{(0)}_{\mu\nu}(q) + \chi^2 g^{(2)}_{\mu\nu}(q)+\cdots
\eqx
where $g^{(0)}_{\mu\nu}$ is the metric for the boundary system.
In three dimensions, the boundary stress energy tensor is then given by \cite{de Haro:2000xn}
\eq
T_{\mu\nu}(q) = {1\over 8\pi G}
\left[g^{(2)}_{\mu\nu}(q)-g^{(0)}_{\mu\nu}(q)\,g^{(2)}_{\alpha\beta}(q) g^{(0)\alpha\beta}(q)\right]+
\tau_{\mu\nu}(q) 
\eqx
where $\tau_{\mu\nu}(q)$ is the scalar contribution for the stress energy tensor
given by
\eq
\tau_{\mu\nu}(q)={1\over 8\pi G}\Big[\partial_\mu \phi_B\partial_\nu \phi_B
- {g^{(0)}_{\mu\nu}\over 2}\, g^{(0)\alpha\beta}\partial_\alpha \phi_B\partial_\beta \phi_B \Big] 
\eqx
with $\phi_B$ denoting the boundary value of the scalar field.
For our case, the boundary metric is given by
\eq
g^{(0)}_{\mu\nu}={\rm diag}(-1,1)=\eta_{\mu\nu} 
\eqx
since the boundary system is defined in the flat Minkowski space
in two dimensions and the scalar contribution to the stress energy tensor vanishes
since the scalar field is constant except $x=0,\,\, l$.
Let us first bring the metric in (\ref{bmetric})
to the form
\eq
ds^2= {dp^2\over \sin^2 p} +{dq_1^2\over \sin^2 p}\Big(1-{\gamma^2\over 4} D_1(x,y) \,\Big)
-{dq_0^2} \cot^2 p\, \Big(1-{\gamma^2\over 4} D_2(x,y)\,\Big)
\eqx
where $q_0=-i\tau$.
Introducing ${\cal X}_{FG}(x,y)$ and ${\cal Y}_{FG}(x,y)$ by
\eqn
q_1(x,y)=x-{{\gamma^2\over 8}} {\cal X}_{FG}(x,y)+O(\gamma^4),
\qquad p(x,y)= y -{{\gamma^2\over 8}} {\cal Y}_{FG}(x,y)+O(\gamma^4) 
\label{fg}
\eqnx
and comparing the two forms of the  metric to the  $O(\gamma^2)$, one finds the
differential equations
\eqn
 \partial_y  {\cal Y}_{FG} -  {\cal Y}_{FG}\,\cot y = a(x,y) ,
 \quad
\partial_x  {\cal Y}_{FG} +\partial_y  {\cal X}_{FG}=0 
\label{fgeq}
\eqnx
with
\eqn
&& D_1(x,y)=a(x,y)- \partial_x  {\cal X}_{FG} +  {\cal Y}_{FG}\, \cot y ,
\nonumber\\
&& D_2(x,y)=b(x,y)+{ {\cal Y}_{FG}\over \sin y\cos y} .
\eqnx
The boundary conditions
$D_1(x,0)=D_2(x,0)=0$
are required to have the standard form of the boundary metric
$\eta_{\mu\nu}$.

For the regions $x \ge l$ and  $x \le 0$, after change of the coordinate transformation in (\ref{coortr}),
the transformed solutions behave
\eq
a'(x',y')=O(\sin^4 y')\,, \quad\quad\quad    b'(x',y')=O(\sin^4 y')
\label{fgxy0}
\eqx
and one can show that there is no stress-energy contribution except that of the
original, undeformed  BTZ black hole.  

Let us turn our attention to the  region between the interfaces, $0 \le x \le l$. For the diagonal part, 
the asymptotic behaviors in (\ref{fgxy0}) still hold in this region and, thus, it 
does not contribute to the 
stress tensor.
For the case of the cross-term,  the situation is different  and one  expects an extra
contribution to the stress tensor, which leads to a nonvanishing Casimir energy.
 We shall analyze the details  by constructing
the corresponding Fefferman-Graham metric in the near-boundary  region.

We note that the solution has the expansion
\eqn
&& a(x,y) =a_1(x)\sin^2 y +a_2(x)\sin^4 y +O(\sin^6y) ,\\
&& b (x,y) =b_0(x) +b_1(x)\sin^2 y +O(\sin^4y)
\eqnx
in the region between the interface.
Then the corresponding solution of satisfying (\ref{fgeq}) the boundary condition $D_1(x,0)=D_2(x,0)=0$
is given by
\eqn
\label{fgx}
&& {\cal X}_{FG}=  -\int dx\, b_0(x) + b'_0(x)  (1-\cos y) +O(\sin^4 y) ,\\ 
&&{\cal Y}_{FG}= \left[ -b_0(x) +a_1(x) (1-\cos y) +O(\sin^4 y)\,\, \right] \sin y ,
\eqnx
where the prime denotes the derivative with respect to the argument.
Then $D_2(x,y)$ and $D_2(x,y)$ can be expanded as
\eqn
&& D_1(x,y)= -\f{b''_0-b_0-3a_1}{2}\sin^2 y +O(\sin^4 y) ,\\
&& D_2(x,y)= \f{b''_0-b_0-3a_1}{2}\sin^2 y +O(\sin^4 y) ,
\eqnx
where we have the relation
\eq
b_1(x)=-\f{4a_1-b''_0}{2}
\eqx
which holds for any nonsingular solutions of the geometric part.

Now explicitly,
\eqn
&&a_1(x)=
-2  \f{1}{\sinh x \sinh (l-x)} \left( {\cal G}(m_0)+{1\over 2}\Delta(l) G_h(m_0) \right) ,\\
&&b_0(x)=
-2  \f{4} { m_0+ \f{1}{m_0} +2 \cosh l  } \left( {\cal G}(m_0,\cosh l)+{1\over 2}\Delta(l) G_h(m_0) \right) ,
\eqnx
where
\eq
 m_0= \f{\sinh (l-x)}{\sinh x} .
\label{m0}
\eqx
A crucial simplification comes from the relation
\eq
b''_0-b_0-3a_1=(-2)\f{8}{\sinh^2 l} ,
\eqx
where the extra factor $(-2)$ reflects the fact that the cross term for the current case comes with extra factor $(-2)$.
Thus finally,
\eqn
&& D_1(x,y)= \f{8}{\sinh^2 l}\sin^2 y +O(\sin^4 y) ,\\
&& D_2(x,y)= -\f{8}{\sinh^2 l}\sin^2 y  +O(\sin^4 y) .
\eqnx


The rest is straightforward.
Noting
\eq
\chi=2\tan {q\over 2} 
\eqx
one finds
\eq
g_{\mu\nu}^{(2)}=\left({1\over 2}-\f{2}{\sinh^2 l} \Theta(x)\Theta(l-x)\gamma^2
\right){\rm diag}(+1,+1)+O(\gamma^4) 
\eqx
with the step function $\Theta(x)=\f{1+\epsilon(x)}{2}$.
Therefore, one has
\eq
T_{\mu\nu} ={1\over 16\pi G}\left(1-\f{4\gamma^2}{\sinh^2 l} \Theta(x)\Theta(l-x)
\right) {\rm diag}(+1,+1)+O(\gamma^4) .
\eqx
Finally recovering the temperature dependence by the scaling transformation, we have
\eq
T_{\mu\nu}= \f{\pi^2 \, T^2- \f{4\pi^2 T^2\gamma^2}{\sinh^2 2\pi T l} \Theta(x)\Theta(l-x) }
{4\pi G} {\rm diag}(+1,+1) +O(\gamma^4)
\label{stress}
\eqx
which clearly shows the effect of the Casimir energy between the interfaces.
The Casimir energy density and pressure are respectively given by
\eqn
&& {\cal E}_C = -\f{2c}{3} \f{\pi T^2}{\sinh^2 2\pi T l} \Theta(x)\Theta(l-x)\,  \gamma^2 + O(\gamma^4) , \\
&& p_C= -\f{2c}{3} \f{\pi T^2}{\sinh^2 2\pi T l} \,  \gamma^2 + O(\gamma^4) ,
\label{cpressure}
\eqnx
and the pressure is acting on the interfaces.
The Casimir energy contribution is negative and the corresponding pressure is negative (attractive). For the general case of 
double interfaces,  the factor $-2$ in the above expression is replaced by $2 \alpha_+ \alpha_-$. 
For the double interfaces with $\alpha_+ \alpha_-  > 0$,
the Casimir energy contribution  becomes now positive reflecting the repulsive nature of the same-signature interfaces.
For the general double interfaces, the energy becomes 
\eq
E= \f{c}{6} \left[  \pi T^2 L + \f{ 2\pi T^2 l}{\sinh^2 2\pi T l} (2 \alpha_+\alpha_-)\gamma^2 +O(\gamma^4)
\right]
\label{doubleenergy}
\eqx
where the full size $L$ of the  system should be taken to be large to avoid any finite size effect.

\subsection{Geometric entropy of the double interfaces}
In this section  we shall  compute the entropy of the double-interface system from the geometric side.  
In the later section, we shall compare the results  with those from the direct 
field-theory computation and show that they agree with each other. 

The interface contribution from the diagonal terms  has been 
computed  in \cite{bgj} to all order in $\gamma$.  For each of unit-coefficient diagonal term, the entropy 
correction reads \cite{bgj}
\eq
\delta S^0_{\rm diag} =\f{1}{4G} \gamma^2  +O(\gamma^4) .
\eqx
Therefore, the total diagonal contribution for the double interfaces reads
\eq
\delta S^{\rm diag} =\f{1}{4G}(\alpha_+^2 + \alpha^2_-) \gamma^2  +O(\gamma^4) = \f{1}{2G} \gamma^2  +O(\gamma^4)
\label{diagentropy}
\eqx
with $\alpha_-=-\alpha_+=1$.

Let us compute here the entropy correction due to the off-diagonal
terms. For this we put the boundary system in a box with a size $L=2 w_0$ and consider the region
$-w_0+ \f{l}{2}\le q_1\le w_0+\f{l}{2}$  where  $q_1$ is the Fefferman-Graham coordinate defined in (\ref{fg}). 
To avoid any finite size effect, we shall
assume $w_0 \gg l$ and $w_0 \gg 1$ and consider one half of the box specified by
$\f{l}{2}\le q_1 \le w_0+\f{l}{2}$ utilizing the symmetry of the solution we are considering.

\begin{figure}
\begin{center}
\includegraphics[scale=1.2]{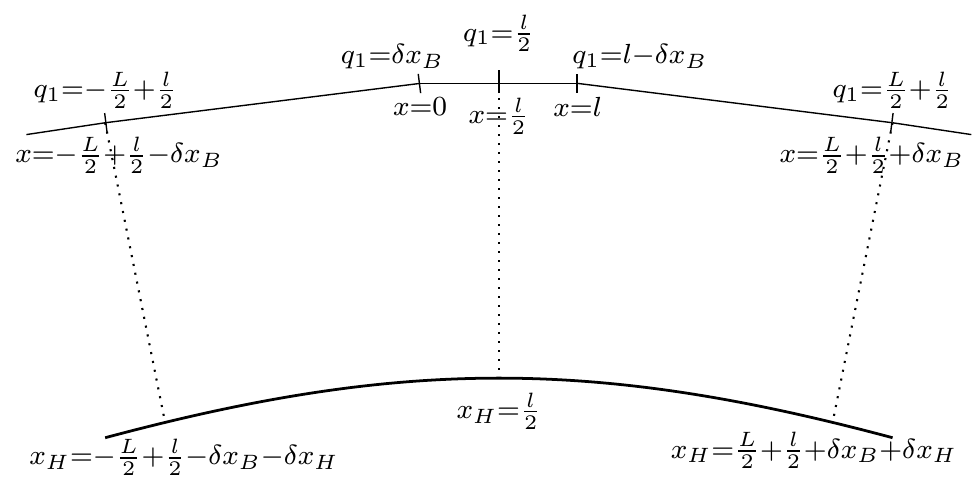}
\end{center}
\caption{ The relation between $x$ and $q_1$ coordinate along the boundary is depicted for the double interfaces 
with the coefficients $\alpha_-=-\alpha_+=1$. The thick line represents the horizon located at $y=\pi/2$. The dotted lines 
represent the trajectories determined by the boundary-horizon map. }
\label{bmap}
\end{figure}

First let us translate the boundary coordinate $q_1$ to our original coordinate $x$. 
As identified in (\ref{fg}), (\ref{fgxy0}) and (\ref{fgx}), 
one has
\eq
q_1 = x +\f{\gamma^2}{8}\Theta(x) \Theta (l-x)  \int^x_{\f{l}{2}} dx' b_0 (x') +O(\gamma^4)
\eqx
which is along the boundary we identified in the previous section.  For $x= \f{l}{2}$, we let $q_1= \f{l}{2}$ by adjusting the
integration constant.  Let $x_B$ be the $x$-coordinate  value corresponding to $q_1= w_0 +\f{l}{2}$. The from the above relation,
one finds
\eq
x_B=  w_0 +\f{l}{2} + \delta x_B 
\eqx
with
\eq
\delta x_B = -\f{\gamma^2}{8}  \int^l_{\f{l}{2}} dx' b_0 (x')+O(\gamma^4) .
\eqx
We evaluate this expression for the unit-coefficient cross term 
\eq
b^c_0(x)= \f{4} { m_0+ \f{1}{m_0} +2 \cosh l  } \left( {\cal G}(m_0,\cosh l)+{1\over 2}\Delta(l) G_h(m_0) \right)
\eqx
where $m_0(x)$ is given  in (\ref{m0}) and we strip off the factor $2\alpha_+ \alpha_-$. Of course, we shall recover this factor in 
order to obtain  the full contribution.
The integral can be straightforwardly evaluated, which leads to
\eq
\delta x_B = -\f{\gamma^2}{2} \left(
\coth l -\f{l}{\sinh^2 l}
\right)+ O(\gamma^4) .
\eqx
The details of computation are relegated to Appendix C.   We  illustrate the relation between $x$ and $q_1$ coordinates
along the boundary in Fig.~\ref{bmap}.

Next we turn to the problem of the boundary-horizon map to find the horizon point corresponding to
$x=x_B$. (For $x=\f{l}{2}$, we choose $x_H= \f{l}{2}$.)  
We shall use the method  \cite{bgj} based on the conserved current. 
(Alternatively one may use the method  based on the lightlike geodesics orthogonal to the boundary \cite{Bhattacharyya:2008xc}, which leads to the 
same result in the large size limit.)
The condition reads
\eq
dx^a \epsilon_{abc}\nabla^b \xi^c=0
\eqx
where $\xi^a$ is the time translation Killing vector.
This leads to 
\eq
{dx\over dy}={g_{yy}\, \partial_x g_{tt} \over g_{xx}\, \partial_y g_{tt}}
={\gamma^2\over 8} \sin y\cos y \, \partial_x b(x,y)+O(\gamma^4)\,.
\eqx
Note that
\eq
b_c(x,y)= 2 \Delta(l) \f{e^{x-\f{l}{2}}}{\sin y} +O(e^{-x})
\eqx
for  $x \gg l$ and $x \gg 1$.
The above equation is solved by
\eqn
 x(y)=x_B
+\f{\gamma^2}{4}\left[  \Delta (l) \sin y \, e^{w_0} +O(e^{-w_0}) \right]
+O(\gamma^4)
\eqnx
which describes the map from the boundary point $x= x_B$ to the corresponding horizon point.
Then at the horizon, the corresponding coordinate $x_H$ is
given by
\eq
x_H=w_0 +\f{l}{2} + \delta x_B + \delta x_H
\eqx
with
\eqn
\delta x_H =\f{\gamma^2}{4}\left[   \Delta(l) e^{w_0} +O(e^{-w_0})\right] .
\eqnx
The horizon length for the entropy then
becomes
\eqn
L_H/2 &=&\int^{x_H}_{\f{l}{2}} dx \Big(1-{\gamma^2\over 8}a(x, {\f{\pi}{2}})\Big)+O(\gamma^4)\,\nonumber\\
&=& (w_0 +\delta x_B+\delta x_H ) -{\gamma^2\over 8}\int^{w_0+\f{l}{2}}_{\f{l}{2}} dx\, a(x, \f{\pi}{2})+O(\gamma^4) .
\eqnx
The integral can be evaluated as
\eqn
&&\int^{w_0+\f{l}{2}}_{\f{l}{2}} dx\, a_c(x, {\pi/2})=
-4 +2 \sqrt{2 \cosh 2 w_0 + 2 \cosh l} \left(
\Delta(l)- 2{\cal I}(e^{-2 w_0},\cosh l)
\right)\nonumber\\
&& \quad\quad\quad = -4 + 2 e^{w_0}\Delta(l) +O(e^{-w_0})
\eqnx

Thus the length correction due to the unit-coefficient cross term becomes
\eq
\delta_c L_H= \gamma^2 \left[1-\Big(
\coth l -\f{l}{\sinh^2 l}
\Big)\right] +O(\gamma^4)
\eqx
in the large $L$ limit.
Therefore, the total change of the entropy for the two interfaces is given by
\eq
\delta S ={1\over 4 G}(L_H-L)={1\over 4G} \left[
(\alpha_+ + \alpha_-)^2- 2 \alpha_+\alpha_-  \Big(
\coth l -\f{l}{\sinh^2 l}
\Big)
\right] \gamma^2 +O(\gamma^4)
\label{linearentropy}
\eqx
where we have added the diagonal  contribution (\ref{diagentropy}).

\begin{figure}
\centering
\includegraphics[scale=1]{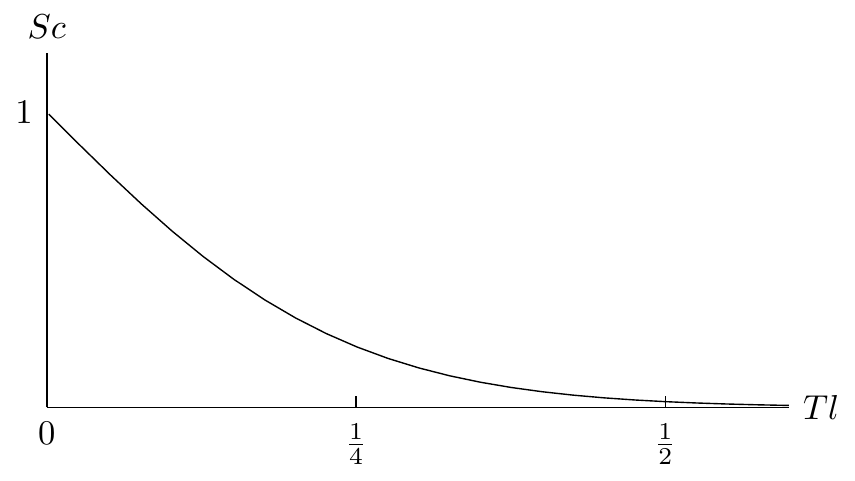}
\caption{ The entropy of the cross term contribution is depicted for $\alpha_-=\alpha_+=1$, $c=3$, and $\gamma^2=1$. It shows that
this contribution of entropy vanishes for the large $Tl$ limit.}
\label{crossentropy}
\end{figure}

Recovering the temperature dependence, one finds
\eq
S=S_0+\delta S =S_0+\f{c}{6} \left[ 
(\alpha_+ + \alpha_-)^2  - 2 \alpha_+\alpha_-  \Big(
\coth 2 \pi T l -\f{ 2\pi T l}{\sinh^2 2\pi T l}
\Big)
\right]\gamma^2+O(\gamma^4)
\label{linearentropy1}
\eqx
where $S_0$ denotes the undeformed BTZ entropy 
\eq
S_0=\f{c}{3} \pi T L .
\eqx
When $\alpha_-= -\alpha_+ =1$, the entropy correction
\eq
\delta S =\f{c}{3} \Big(
\coth 2 \pi T l -\f{ 2\pi T l}{\sinh^2 2\pi T l}
\Big)
\gamma^2+O(\gamma^4)
\eqx
comes from the thermodynamic effect  related to the Casimir energy.  
As $lT\rightarrow 0$, the two interfaces join together and become a single
interface with an interface coefficient $\alpha_+ + \alpha_-$. Indeed the entropy correction approaches
that of the single interface:
\eq
\delta S \ \rightarrow \  \f{c}{6}
(\alpha_+ + \alpha_-)^2 \gamma^2+O(\gamma^4) .
\eqx 
On the other hand, in the limit where $LT$ becomes infinity, one expects that the two interfaces become
independent with each other. The corresponding entropy correction becomes that for the two 
independent interfaces,
\eq
\delta S \ \rightarrow \  \f{c}{6}
(\alpha^2_+ + \alpha^2_-) \gamma^2+O(\gamma^4)
\eqx
 The behavior of the cross-term entropy correction (as a function of $LT$), for $\alpha_-=-\alpha_+=1$ and $c=3$, 
is illustrated in Fig.~\ref{crossentropy}. It clearly shows that the contribution from the cross-term vanishes in the large separation limit
of the interface distance.



\section{Check of the correspondence}

As discussed previously in \cite{Bak:2007jm}, the corresponding field-theory dual to the
Janus deformation is defined by the Lagrangian density\footnote{Strictly speaking, one has to use the boundary coordinate
$q_1$ instead of $x$ since $q_1$ differs from $x$. But for the sake of simplicity of presentation, we shall here use 
$x$ instead of $q_1$ and, hence,  $x$ should be understood as the boundary coordinate $q_1$ in this section.}
\eq
{\cal L}(t,x) ={\cal L}_0 (t,x) \, e^{-\phi_B(x)} 
\eqx
where ${\cal L}_0$ is the Lagrange density for the original, undeformed CFT and $\phi_B$
denotes the boundary value of the bulk scalar field $\phi(x)$ which is dual to the Lagrange density
operator. 
For the deformation of the current problem,
\eq
{\cal L}(t,x) ={\cal L}_0 (t,x) \, \left[
1- \gamma \sum_n \alpha_n \epsilon(x-l_n) +O(\gamma^2)
\right]
\eqx
where the boundary condition (\ref{generalbc}) is used. We shall consider  the
finite-temperature field theory which 
is dual to our deformed  black hole geometry constructed in the above.  

The free energy, 
\eq
F = -\f{1}{\beta} \ln {\rm tr}\, e^{-\beta H}
\eqx
can be computed perturbatively as a power series in $\gamma$ 
using the so-called conformal perturbation theory. Namely,
\eqn
&&{\rm tr}\, e^{-\beta H} \nonumber\\
&& ={\rm tr}\, e^{-\beta H_0}\left[ 
1- \f{1}{2}\int^\beta_0 \! d\tau \int^\beta_0 \! d\tau' \int^\infty_{-\infty}
dx  \int^\infty_{-\infty}
dx' \langle \delta{\cal L}(-i\tau,x)  
\delta {\cal L}(-i\tau',x')\rangle + O(\gamma^4)
\right]
\eqnx
where we have used the fact that
\eq
\langle {\cal L}_0(-i\tau,x)  \rangle =0 .
\eqx
The expectation value is defined with respect to the undeformed (finite-temperature)  CFT in the regime 
where the gravity and  the CFT correspondence is valid. See \cite{Bak:2007jm} for the details of the conditions
for the gravity approximation.
The free energy  can be expanded as
\eq
F= F_0 +  \gamma^2 F_2+ O(\gamma^4)
\eqx
where $F_0$ is the undeformed BTZ free energy given by 
\eq
F_0= -\f{c}{6} \pi T^2 L .
\eqx
Using the two-point function \cite{Bak:2007qw} given by
\eq
\langle {\cal L}(-i\tau,x) {\cal L}(-i\tau',x')\rangle = \f{1}{16 \pi^2 G}\f{(2\pi T)^4}{\big[\cos(\f{2\pi (\tau-\tau')}{\beta} )-
\cosh(\f{2 \pi(x-x')}{\beta})-i\epsilon\big]^2} 
\eqx
and performing the $\tau$ and $\tau'$ integrals, one finds $F_2$ is given by
\eq
F_2 =-\f{T}{8 G} \sum_{n,m}\alpha_n\alpha_m
\int_{-\infty }^{\infty}dx \epsilon(x- 2\pi T l_n)\int^{\infty}_{-\infty} dx' \epsilon(x'-2\pi T l_m)\,
h(x-x')
\eqx
where we rescaled $x$ and $x'$ by the factor $2\pi T$ and $h(x)$ is given by
\eq
h(x)=\f{\cosh(x)+q^2}{\big((\cosh(x)+q^2\big)^2-1)^{3/2}}
\eqx
with $q^2=i\epsilon$.
(The above expression  is for the case of infinite-sized box and has to be regulated appropriately  by putting the system in a box
with a  finite size $L$.)   
For the further evaluation, let us first consider the unit-coefficient cross-term
\eq
F^c_2 =-\f{T}{8 G} 
\int_{-\infty }^{\infty}dx \epsilon(x)\int^{\infty}_{-\infty} dx' \epsilon(x'-2\pi T l)\, h(x-x') .
\eqx
Noting $h(x)= h(-x)$, the above integral is rearranged as
\eqn
F_2^c &=& -\f{T}{8 G} 
\int_{0 }^{\infty}dx \int^{\infty}_{-\infty} dx' \left(\epsilon(x'-2\pi T l) +\epsilon(x'+2\pi T l)
\right) h(x-x') \nonumber\\
&=&  -\f{T}{4 G} 
\int_{0 }^{\infty}dx \int^{\infty}_{2\pi Tl} dx' \left(   h(x-x')-  h(x+x') \right)\nonumber\\
&=& - \f{T}{2 G} \left[
W_1 (\infty) - W_2(2\pi Tl)
\right]
\eqnx
where 
\eqn
W_1(w) &=& \f{1}{2} 
\int_{0 }^{w}dx \int^{\infty}_{0} dx' \left(   h(x-x')-  h(x+x') \right)\nonumber\\
&=& \int_{0 }^{w}dx \int^{x}_{0} dx' \, h(x')
\eqnx
and
\eqn
W_2(w) &=& \f{1}{2} 
\int_{0 }^{\infty}dx \int^{w}_{0} dx' \left(   h(x-x')-  h(x+x') \right)\nonumber\\
&=& \int_{0 }^{w}dx' \int^{x'}_{0} dx \, h(x) .
\eqnx
For the last equality in the above, we have interchanged the order of integrations on $x$ and $x'$. $W_2(2\pi Tl)$ 
does not require any 
regularization related to the box size of the system. On the other hand, $W_1(\infty)$ requires a regularization
and we shall change the range of integral $(-\infty, \infty)$ to $[-\pi T L, \pi T L]$ leading to 
the regularized expression $W_1(\pi TL)$. Later we shall take the large $TL$ limit to avoid any finite size effect related to the
size of the box. 
Note that $W_1(w)=W_2(w)= W(w)$. The integral $W(w)$ is found in \cite{bgj} and the result reads
\eq
W(w)=\f{w}{2q^2} -\f{1}{\sqrt{2}\,q} +\f{1}{2} \coth w + O(q^2)
\eqx
and, for the regularization with respect to $q^2(=i\epsilon)$, we shall keep only the finite part, i.e.
$\f{1}{2} \coth w$. Therefore,
\eq
F_2^{\rm cross} =-\f{T}{4G} \big(1-\coth 2\pi Tl \big)
\eqx
where we take the large $TL$ limit ignoring the finite size effect related to the box size.

For the case of a unit-coefficient  diagonal term, one can follow the same procedure leading to the result
\eq
F_2^{\rm diag}= -\f{T}{4G} . 
\eqx
Therefore, for the general double interface,  one is led to
\eq
F= -\f{c}{6}\left[
\pi T^2 L + T \Big((\alpha_+ + \alpha_-)^2 
- 2\alpha_+  \alpha_-  \coth 2\pi Tl 
\Big) \gamma^2 + O(\gamma^4) .
\right]
\eqx
Using the standard thermodynamic relations,
\eq
S= -\f{\partial F}{\partial T}\,, \ \ \ E=F-TS
\eqx 
one finds the precise agreement with those from geometric side given in (\ref{doubleenergy})and (\ref{linearentropy}).
One can check the first law
\eq
TdS =dE + p_L dL + p_l dl 
\eqx
where 
\eq
p_L = -\f{\partial F}{\partial L} =\f{c}{6} \pi T^2, \quad  p_l = -\f{\partial F}{\partial l}=
\alpha_+ \alpha_- \, \f{c}{3}  \f{ 2\pi T^2}{\sinh^2 2\pi T l}\gamma^2 + O(\gamma^4)
\eqx
One can check that $p_l$ agrees with $p_c$ in (\ref{cpressure}) from the geometric side  with $\alpha_+\alpha_-=-1$. Of course 
$p_L$ is in agreement with that from
the stree energy tensor in (\ref{stress}).

\section{General interfaces and the lattice}
Until now, we have discussed the detailed properties of the double interface configurations. 
Extending this analysis
 to the general interface system is rather straightforward. 
In this section, we shall describe the  boundary  shape and 
the thermodynamic properties of  the general interface system.
We shall also comment on  the properties of the 
interface lattice briefly.

\begin{figure}
\centering
\includegraphics[scale=1.]{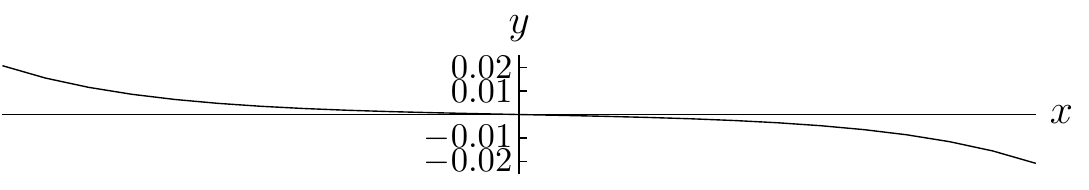}
\caption{The trajectory $y'=0$  is depicted  in $(x,y)$ space for  the coordinate transformation in relation with the 
homogeneous solution $X=\f{\sinh x}{\sin y}$. We take $l=1$ and  $\gamma^2=0.01$.}
\label{bh5}
\end{figure}

The solution for the general interface configuration is constructed in (\ref{generalbc}),
(\ref{gscalar}), (\ref{gdiag}) and (\ref{offdiag}).  Finding the boundary shape  for a given 
interface solution is rather involved. The boundary shape in a particular coordinate system  is dependent on the choice of the 
homogeneous terms though the geometry itself including its boundary should be independent of this choice.   
This is because different choices of homogeneous solutions are all related by  coordinate transformations, as was verified
in the previous section. This implies that the shape of the boundary in one coordinate system is affected by the choice 
of the homogeneous terms.  For the homogeneous term $X=\f{\sinh x}{\sin y}$ that is related to the diagonal solution $q_0(x)$, 
the part of coordinate transformation in (\ref{coortr}) takes the form
\eq
y' = y + \f{\gamma^2}{8} \cos y \sinh x +O(\gamma^4) .
\eqx
The trajectory of $y'=0$ is described by
\eq
\sin y= - \f{\gamma^2}{8} \sinh x +O(\gamma^4) .
\eqx
This is illustrated in Fig.~\ref{bh5} for $\gamma^2=0.01$ and $l=1$.

\begin{figure}
\centering
\includegraphics[scale=1]{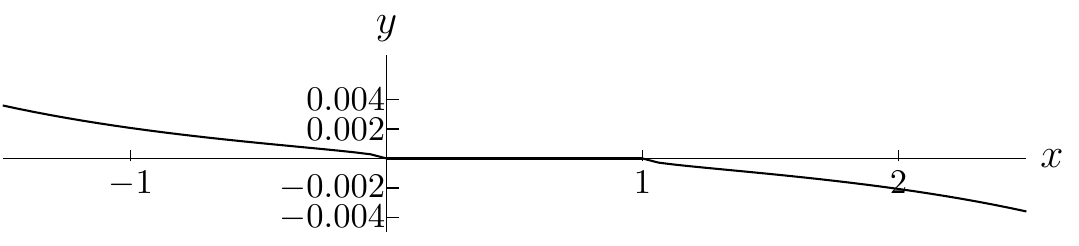}
\caption{The trajectory $y'=0$  is depicted  in $(x,y)$ space for  the coordinate transformation in relation with the 
homogeneous solution $G_h(m)$. We take $l=1$ and $\gamma^2=0.01$. }
\label{bh6}
\end{figure}

It is rather clear that the coordinate transformation is highly nonlinear, which means 
that any particular  trajectory, e.g. $y'=0$, has different-looking shapes depending the  choices of  coordinate system. 
Similarly for the homogeneous term $G_h(x,y,\cosh l)$ related to the cross-term  part of the solution,  the trajectory
$y'=0$ has an even complicated shape in the $(x,y)$ coordinate system. Using (\ref{coortr}), one can identify 
the trajectory $y'=0$ as
\eq
\sin y =\f{\gamma^2}{8}\left[ \Theta (-x)\sqrt{\sinh (-x) \sinh (l-x)} -\Theta (x-l)\sqrt{\sinh x \sinh (x-l)}\,\right] 
+O(\gamma^4) .
\eqx 
As we see in Fig.~\ref{bh6}, the trajectory $y'=0$ in the $(x,y)$ plane reflects the highly nonlinear nature of the coordinate transformation.
Hence depending on the choice of  homogeneous terms, the shapes of the boundary in the $(x, y)$ plane have to change rather 
drastically
and one cannot expect any kinds of shape invariance of  
under the coordinate transformations.

\begin{figure}
\centering
\includegraphics[scale=1.3]{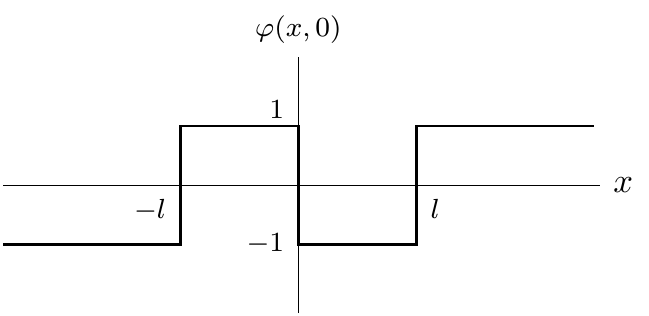}
\caption{The boundary condition, $\varphi(x,0)$,  is depicted for the three interface configuration we consider in (\ref{threei}). }
\label{bh4}
\end{figure}

Having this complication in mind, with a specific  choice of the homogeneous terms, one can in general find the 
boundary  shape of the geometry for the general interface system. For the cases of double interfaces, we have illustrated these 
boundary shapes in Figs.~\ref{boundaryfig} and \ref{boundaryfigb}. As a further illustration, here we consider a three interface 
configuration  with a boundary condition of the scalar field,
\eq
\varphi(x,0)= \epsilon(x+l) -\epsilon(x) +\epsilon(x-l) .
\label{threei}
\eqx
We illustrate this boundary condition in Fig.~\ref{bh4}. 

\begin{figure}
\centering
\includegraphics[scale=1.2]{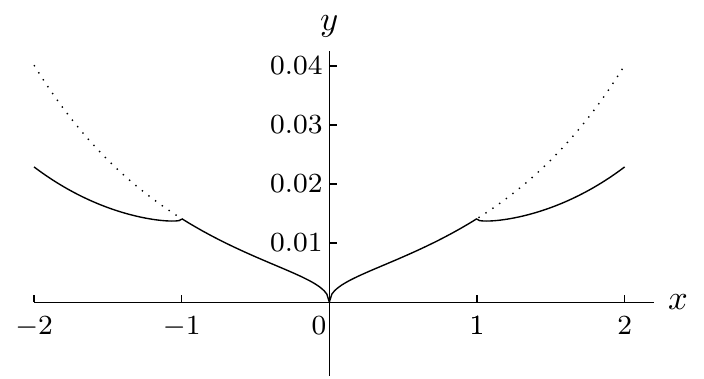}
\caption{The shape of the boundary (solid line) in $(x,y)$ space is depicted for  the three interface system with $l=1$ and $\gamma^2=0.01$. 
The three cusps represent the locations of
the interfaces. The dotted lines represent the function in (\ref{bshape-re1}) for $x > l$ together with  its image under the reflection with respect to $y$-axis. }
\label{bh7}
\end{figure}

The corresponding geometric part  to the  $O(\gamma^2)$ is described by 
\eqn
a(x,y)&=&q_0(x+l)+\frac{1}{2} \Big( q_0(x)+q_0(-x) \Big) +q_0(l-x)
\nonumber\\
&&-2 \Big(a_c^0(x,y,l) +a_c^0(x+l,y,l) \Big) +2 a_c^0(x+l,y,2l) ,\\
 b(x,y)&=&q_0(x+l)+\frac{1}{2} \Big( q_0(x)+q_0(-x) \Big) +q_0(l-x) \nonumber\\
&&  -2 \Big(b_c^0(x,y,l) +b_c^0(x+l,y,l) \Big) +2 b_c^0(x+l,y,2l) .
\eqnx
The homogeneous part of the solution is chosen in such a way that the geometric part of the
solution has a symmetry under the exchange $x\leftrightarrow -x$. For $0 \le x \le l$,
the boundary is described by
\eq
\sin y =\gamma^2  \left[
  \Delta (l) \sqrt{\sinh x \sinh (x+l)} -\f{3\pi}{16} \sinh x 
\right]+O(\gamma^4)
\label{bshape-re1} 
\eqx
while, for $l \le x $,
\eqn
\sin y &=& \gamma^2  \left[
  \Delta (l) \sqrt{\sinh x \sinh (x+l)} -\f{3\pi}{16} \sinh x -\f{3\pi}{8} \sinh (x-l) 
\right. \nonumber\\
&&\left. \quad +\Delta (l) \sqrt{\sinh x \sinh (x-l)} -\Delta (2l) \sqrt{\sinh (x+l) \sinh (x-l)}\,\right]+O(\gamma^4) .
\label{bshape-re2} 
\eqnx
The corresponding shape of the boundary in $(x,y)$ plane  is illustrated in Fig.~\ref{bh7}.

Next we turn to the problem of the thermodynamics of general interface system. 
The analysis for the double interface configuration can be extended immediately  by
the same analysis. The free energy can be identified as
\eq
\delta F= -\frac{c}{6}\left[ T \sum_n \alpha_n^2 +2T \sum_{m < n}\alpha_m\alpha_n (1- \coth 2\pi T l_{mn})
\right]\gamma^2 + O(\gamma^4)
\eqx
where $\delta F= F- F_0$ and  $l_{mn}= l_n-l_m$.
It  follows that the energy and the entropy corrections become
\eqn
 \delta E &=&\frac{c}{3}  \sum_{m < n}\alpha_m\alpha_n \f{2\pi T^2 l_{mn}}{ \sinh^2 2\pi T l_{mn}}
\gamma^2 + O(\gamma^4) ,\nonumber\\
 \delta S &=&\frac{c}{6}\left[ \sum_n \alpha_n^2 +2 \sum_{m < n}\alpha_m\alpha_n 
\Big(1- \coth 2\pi T l_{mn}+
\f{2\pi T l_{mn}}{ \sinh^2 2\pi T l_{mn}}
\Big) \right]
\gamma^2 + O(\gamma^4) .
\eqnx
One can verify the first law of the thermodynamics
\eq
T dS = dE + p_L dL +\sum_n p_{l_n} dl_n
\eqx
where 
\eq
p_{l_n}=-\f{\partial F}{\partial \, l_n}=\f{c}{3}\alpha_n \left[
\sum_{m<n} \alpha_m \f{ 2\pi T^2}{\sinh^2 2\pi T l_{mn}} -\sum_{n<m} \alpha_m \f{ 2\pi T^2}{\sinh^2 2\pi T l_{mn}}
\right]\gamma^2 + O(\gamma^4).
\eqx

For the lattice configuration, these corrections diverge due to the invariance under the  translation by $2l$ where $2l$ corresponds
to the lattice spacing.  We instead define thermodynamic quantities per each cell covering an interval of  size $2l$. The energy correction 
 per unit cell 
reads
\eq
\delta E_{2l}= -\f{2c}{3} \sum^\infty_{n=1} (-1)^{n+1}\f{2\pi  nT^2  l}{\sinh^2 2\pi nT  l}\gamma^2 +O(\gamma^4)
\eqx
while the entropy correction per unit cell becomes
\eq
\delta S_{2l}= \f{c}{3} \left[1-2 \sum^\infty_{n=1} (-1)^{n+1}\left(1-\coth 2\pi nT l +\f{2\pi n T  l}{\sinh^2 2\pi nT  l}\right)\right]\gamma^2 +O(\gamma^4) .
\eqx
One can check the convergence of these expressions.

\section{ Interfaces in the $T\rightarrow 0$ limit}
The zero-temperature limit of the interfaces system is defined by taking  
the limit $T\rightarrow 0$.  For any  quantities (in the gravity side) that have the dimension of length, 
one has to multiply $2\pi T$ once we recover the temperature dependence.  The zero temperature 
limit of the solution can be obtained by the $T\rightarrow 0$ limit. Practically we make the following replacement
in the solution:
\eqn
\sinh x \rightarrow x,\   \sin y \rightarrow y, \ \sinh (x-l) \rightarrow (x-l), \ \sinh l \rightarrow l , \  \cosh l \rightarrow 1
\eqnx
and
\eq
X \rightarrow X_0 = \f{x}{y}, \quad \ Y \rightarrow Y_0 = \f{x-l}{y}, \quad m(X,Y)\rightarrow m(X_0,Y_0) .
\eqx
The integral ${\cal I}(m,\cosh 0)$ in the cross-term can be analytically evaluated as
\eq
{\cal I}_0(m) ={\cal I}(m,1)=-\f{\sqrt{m}(3+5m)}{4(1+m)^2} +\f{3}{4} \tan^{-1}\sqrt{m}
\eqx
and the homogeneous solution $G_h(m,\cosh l)$ becomes
\eq
G_h(m,\cosh l)\rightarrow G^0_{h}(m)=\f{(1-m^2)(1+m)}{m\sqrt{m}} .
\eqx
With these replacements, the zero-temperature limit can be found without any complication.

In the zero-temperature limit, the free energy $F$ agrees with the energy $E$:
\eq
F=E=\frac{c}{6}  \sum_{m < n} \f{\alpha_m\alpha_n}{ \pi  l_{mn}} \gamma^2 +O(\gamma^4)  .
\eqx
The entropy of the system becomes
\eq
S= \f{c}{6}\, \Big[\sum_n \alpha_n \, \Big]^2 \gamma^2 +O(\gamma^4) .
\eqx
Hence in this zero temperature limit, the entropy contribution  is the same as the single interface
whose interface coefficient is given by the sum $\sum_n \alpha_n$. Finally, for the lattice,  the Casimir energy per 
unit cell in the zero-temperature
limit can be explicitly summed up as
\eqn
E_{2l}= -\f{c}{3} \sum^\infty_{n=1} \f{(-1)^{n+1}}{\pi  n \, l}\gamma^2 +O(\gamma^4)= -\f{\ln 2}{3\pi }\,\f{c}{l}\,\gamma^2 +  O(\gamma^4) .
\eqnx
 Thus the limit and the construction of  solution is  straightforward in a sense. 

\section{Conclusions}

In this paper, we have constructed the black brane solutions which are dual to the multiple interface field theories.
We have computed their thermodynamic quantities from the gravity and the field theory sides, and found a precise 
agreement. 

In their finite temperature  entropy, it is found that  there are two distinct kinds of corrections 
due to  the interfaces. One is the diagonal contribution which is scale invariant and temperature independent.
The other  is the cross-term contribution that comes from any pairs of two separated interfaces.
It is interesting  in the sense that  it represents a correlated entropy between two separate interfaces.
As expected, this entropy is monotonically decreasing as a function of $T l$ and vanishes 
in the large $T l$
limit\footnote{One observation is that its functional form  is the same as the radial shape of the
BPS monopole. 
We wonder if there are  any reasons behind this coincidence.}. 
This contribution alone can be either positive or negative depending on the signatures of the relevant
interface coefficients but the total entropy of the interfaces should be positive definite.

Not to mention, it is of interest to  find the higher order solutions, which involve contributions from interfaces 
whose number is equal to or larger than three.    Especially the computation of $O(\gamma^4)$ geometric part will 
be interesting together with
improving the field theory computation to the same order.

Finally let us comment on the interface lattice solution that is dual to the interface lattice system.
Note that our lattice solution  in the above does not show any manifest lattice translational symmetry.
Of course this geometry itself has the lattice translational symmetry up to the coordinate 
transformation.  
It is possible to compactify the lattice solution  on a circle whose circumference has the length of  
a lattice cell size.  Its zero temperature limit is different from the global Janus solution in many respects.
For instance, the compactified one involves a Casimir energy of $O(\gamma^2)$ while 
the global Janus  carries only the Casimir energy that is independent of $\gamma$. 
This is rather similar to
the fact that the zero temperature limit of the compactified BTZ solution on a circle does not agree with
the global AdS$_3$. In addition the global Janus solution cannot be decompactified since there
will be a conical singularity at the spatial origin by doing so. Further studies in this direction are 
required.

\section*{Acknowledgement}
We would like to thank Romuald Janik for the collaboration at the initial stage of this work.
DB was 
supported in part by  NRF Mid-career Researcher
Program 2011-0013228. 
HM was supported in part by NRF 2010-0011223.

\appendix

\section{Analytic expression for ${\cal I}(x,\cosh l)$}

\begin{figure}
\centering
\includegraphics[scale=0.7]{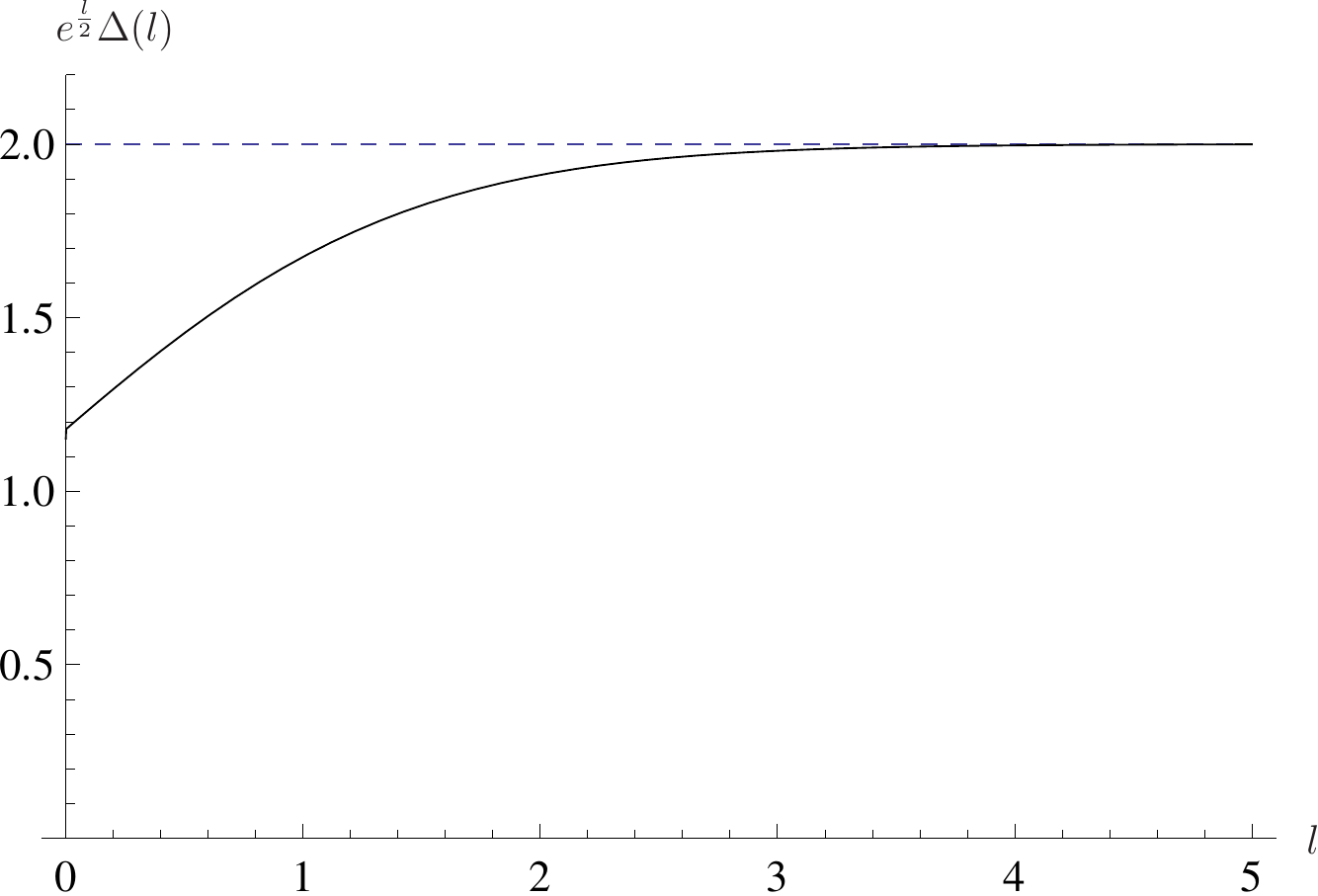}
\caption{The function $\Delta(l) \, e^{\f{l}{2}}$ is depicted. It increases monotonically as a function of $l$ from 
$\f{3\pi}{8}$ to $2$ for the interval  $l\in [0,\infty)$.}
\label{deltaplot}
\end{figure}

The integral plays an important role in our study. Here we present
the analytic expression for this integral. It reads
\eqn
&& {\cal I}(x,\cosh l)=\frac{e^{ l\over 2}}{\sinh^2 l} \left(
\cosh l\  E(\tan^{-1} \sqrt{xe^l}, 1-e^{-2l})-e^{-l} F(\tan^{-1} \sqrt{xe^l},1-e^{-2l})\right.\nonumber\\
&&\quad\quad\quad\quad\quad\quad   \left.- \frac{\sinh l \sqrt{x e^{l}}}
{\sqrt{1+x^2+2x \cosh l}}
\right) ,
\eqnx
where $F(x,k^2)$ and $E(x,k^2)$ are the elliptic integrals of the first and the second kinds.
Then $ {\cal I}(\infty, \cosh l)$ has the expression
\eq
\Delta(l)\equiv {\cal I} (\infty, \cosh l) =\frac{e^{ l\over 2}}{\sinh^2 l }
 \left(
\cosh l \,\,  E( 1-e^{-2l})
-e^{-l}  K(1-e^{-2l})
\right) .
\eqx
One finds
\eq
 {\cal I} (\infty,\cosh 0)={3\pi\over 8}
\eqx
while
\eq
{\cal I} (\infty, \cosh l) \rightarrow 2 e^{-{l\over 2}}  \quad {\rm as} \quad  l \rightarrow \infty .
\eqx
We draw the function $\Delta(l) \, e^{\f{l}{2}}$ in Fig.~\ref{deltaplot}.

For ${\cal I}(x, -\cosh l)$, let us restrict  the argument in the range $0\le x < e^{-l}$. One then get
\eqn
&& {\cal I}(x,-\cosh l)=\frac{1}{\sinh^2 l} \left(
e^{l\over 2}(\sinh l\  F(\sin^{-1} \sqrt{xe^l}, e^{-2l})-\cosh l \ E(\sin^{-1} \sqrt{xe^l},e^{-2l}))\right.\nonumber\\
&&\quad\quad\quad\quad\quad\quad   \left.+ \frac{\sqrt{x}(1-x \cosh l)}
{\sqrt{1+x^2-2x \cosh l}}
\right) .
\eqnx

\section{Relation between $a^c_h(x,y)$ and $b^c_h(x,y)$}
In this appendix, we show the relation
\eq
a^c_h =b^c_h \sin^2 y -\sin y \cos y \, \partial_x b^c_h
\label{relationh}
\eqx
for the homogeneous part of the cross-term solution
\eqn
&&
a_h^c(x,y)=-\frac{
(m-m^{-1})\, \sqrt{{R}(m)}
}{\sqrt{1+X^2}\sqrt{1+Y^2}} ,
\label{ahc}
\\
&&
b_h^c(x,y)=- \f{4(m-m^{-1})
}{\sqrt{R(m)}}
\label{relb}
\eqnx
with
\eq
R(m)\equiv m+m^{-1}+2 \cosh l .
\eqx
Let us note first
\eq
\partial_x \ln m = -\f{\cot y}{2}\frac{
m-m^{-1}
}{\sqrt{1+X^2}\sqrt{1+Y^2}} .
\eqx
Then
\eq
 \sqrt{1+X^2}\sqrt{1+Y^2}\, \partial_x b^c_h =  \f{m-m^{-1}}{R^{\f{3}{2}}(m)}\left[
2\big(m+m^{-1} \big)R(m)-\big(m-m^{-1}\big)^2
\right] \cot y .
\eqx
Inserting this into (\ref{relationh}) with (\ref{relb}), one gets
\eqn
 \sqrt{1+X^2}\sqrt{1+Y^2}\,  a^c_h &=&-\f{m-m^{-1}}{R^{\f{3}{2}}(m)} \left[ \
4 \sin^2 y  \sqrt{1+X^2}\sqrt{1+Y^2}  \, R(m)\right. \nonumber\\
&& \left. +\cos^2 y \Big(
2 (m+m^{-1} )R(m)-(m-m^{-1})^2
\Big)
\right].
\label{arel}
\eqnx
Now using the following identity,
\eq
X^2+Y^2 = 2 XY \cosh l + \f{\sinh^2 l}{\sin^2 y}
\eqx
one can show
\eqn
&&4XY\, R(m) =(m-m^{-1})^2 -4\f{\sinh^2 l}{\sin^2 y} ,\\
&& 4\sqrt{1+X^2}\sqrt{1+Y^2}\,R(m)=2 R (m)(m+m^{-1})-(m-m^{-1})^2 +4\f{\sinh^2 l}{\sin^2 y} .
\label{x1y1}
\eqnx
Inserting (\ref{x1y1}) into  (\ref{arel}), one recovers the expression for $a^c_h$ in (\ref{ahc}).

\section{Evaluation of the integral for $\delta x_B$}
In this appendix, we shall evaluate the integral 
\eq
U(l) \equiv \int^l_{\f{l}{2}} dx\,  b^c_0 (x)
\eqx 
where
\eq
b_0^c(x)= \f{4}{R(m_0)} \left[ 2 m_0 - G_h(m_0)\Big({\cal I}(m_0)-\f{\Delta(l)}{2}\Big)\right]
\eqx
with $m_0$  given in (\ref{m0}).
First note that
\eq
m'_0(x) = -\f{\sinh l}{\sinh^2 x}= - \f{m_0 R(m_0)}{\sinh l} .
\eqx
Using the change of variable, the integral is rearranged as
\eq
U(l)= 4\sinh l \int_0^1 \f{d m_0}{R^2(m_0)} \left(2 
+(1-m^{-2}_0 )R^{\f{1}{2}}(m_0) \Big({\cal I}(m_0)-\f{\Delta(l)}{2}\Big)\right)  .
\eqx
We use the mathematical identity
\eq
\f{1- m^{-2}_0}{R^{\f{3}{2}}(m_0)}  \Big({\cal I}(m_0)-\f{\Delta(l)}{2}\Big) = -\f{2}{R^{2}(m_0)} - \f{d}{d m_0} \f{1}{R^{\f{1}{2}}(m_0)} 
\Big({\cal I}(m_0)-\f{\Delta(l)}{2}\Big) .
\eqx
The integral becomes
\eq
U(l)= 4\sinh l \left[ 4 \int^1_0 \f{dm_0}{R^2(m_0)} - \f{\Big({\cal I}(1)-\f{\Delta(l)}{2}\Big) }{\cosh \f{l}{2}}\right] .
\eqx
We note that
\eq
{\cal I}(1)-\f{\Delta(l)}{2}=- \f{1}{2\cosh \f{l}{2} }
\eqx
and 
\eq
 4 \int^1_0 \f{dm_0}{R^2(m_0)}= \f {\sinh l -l}{\sinh^3 l} .
\eqx
Therefore, one is led to
\eq
U(l)= 4 \left(\coth l - \f {l}{\sinh^2l} \right) .
\eqx

\end{document}